\documentclass[fleqn,usenatbib]{mnras}

\usepackage{newtxtext,newtxmath}

\usepackage[T1]{fontenc}
\usepackage{ae,aecompl}

\usepackage{graphicx}	
\usepackage{amsmath}	
\usepackage{amssymb}	

\newcommand{\kms}{{km~s$^{-1}$}}
\newcommand{\lco}{{L$^\prime_\textrm{CO}$}}

\newcommand{\lir}{L$_\textrm{IR}$}
\newcommand{\alphaco}{{$\alpha_\textrm{CO}$}}

\newcommand{\oiii}{[O~{\sc iii}]}
\newcommand{\Mgas}{M$_\textrm{gas}$}
\newcommand{\deltaMS}{$\Delta_{\textrm{MS}}$}

\title[Impact of quasars on molecular gas]{High molecular gas content and star formation rates in local galaxies that host quasars, outflows and jets}

\author[M.\,E.\,Jarvis et al.]{M.\,E.\,Jarvis,$^{1,2,3}$\thanks{E-mail: miranda.jarvis@gmail.com}
C.\,M.\,Harrison,$^{4}$\thanks{E-mail: christopher.harrison@newcastle.ac.uk} 
V.\,Mainieri,$^{2}$ 
G.\,Calistro Rivera,$^{2}$ 
P.\,Jethwa,$^{5}$ \newauthor  
Z.-Y.\,Zhang,$^{6}$ 
D.\,M.\,Alexander,$^{7}$
C.\,Circosta,$^{8}$
T.\,Costa,$^{1}$
C.\,De Breuck,$^{2}$ 
D.\,Kakkad,$^{9}$\newauthor
P.\,Kharb,$^{10}$
G.\,B.\,Lansbury,$^{2}$ 
A.\,P.\,Thomson$^{11}$
\\
$^{1}$Max-Planck Institut f\"ur Astrophysik, Karl-Schwarzschild-Str. 1, 85748 Garching, Germany \\
$^{2}$European Southern Observatory, Karl-Schwarzschild-Str. 2, 85748 Garching, Germany \\
$^{3}$Ludwig Maximilian Universit\"at, Professor-Huber-Platz 2, 80539 Munich, Germany\\
$^{4}$School of Mathematics, Statistics and Physics, Newcastle University, NE1 7RU, UK \\
$^{5}$Universit\"{a}t Wien, Institut f\"{u}r Astrophysik, T\"{u}rkenschanz-Str. 17, 1180 Vienna, Austria \\
$^{6}$School of Astronomy and Space Science, Xianlin Campus, Nanjing University, 163 Xianlin Avenue, Nanjing, Jiangsu, 210093, P.R.China\\
$^{7}$Centre for Extragalactic Astronomy, Department of Physics, Durham University, South Road, Durham DH1 3LE, UK\\
$^{8}$Dept of Physics \& Astronomy, University College London, Gower Street, London, WC1E 6BT, UK\\
$^{9}$European Southern Observatory, Alonso de Cordova, 3107, Vitacura Casilla 19001, Santiago, Chile\\
$^{10}$National Centre for Radio Astrophysics - Tata Institute of Fundamental Research, Pune University Campus,\\ 
Post Bag 3, Ganeshkhind, Pune 411007, India\\ 
$^{11}$Jodrell Bank Centre for Astrophysics, Department of Physics \& Astronomy, The Alan Turing Building, Upper Brook Street,\\
Manchester M13 9PL, UK\\
}

\pubyear{2020}

\begin{document}
\label{firstpage}
\pagerange{\pageref{firstpage}--\pageref{lastpage}}
\maketitle

\begin{abstract}
We use a sample of powerful $z$$\approx$$0.1$ type~2 quasars (`obscured'; $\log [L_{\text{AGN}}$/erg\,s$^{-1}]$$\gtrsim$$45$), which host kiloparsec-scale ionized outflows and jets, to identify possible signatures of AGN feedback on the total molecular gas reservoirs of their host galaxies. Specifically, we present Atacama Pathfinder EXperiment (APEX) observations of the CO(2--1) transition for nine sources and the CO(6--5) for a subset of three. We find that the majority of our sample reside in starburst galaxies (average specific star formation rates of 1.7~Gyr$^{-1}$), with the seven CO-detected quasars also having large molecular gas reservoirs (average \Mgas=1.3$\times$10$^{10}$~M$_\odot$), even though we had no pre-selection on the star formation or molecular gas properties. Despite the presence of quasars and outflows, we find that the molecular gas fractions (\Mgas/M$_\star$=0.1--1.2) and depletion times (\Mgas/SFR=0.16--0.95~Gyr) are consistent with those expected for the overall galaxy population with matched stellar masses and specific star formation rates. Furthermore, for at least two of the three targets with the required measurements, the CO(6--5)/CO(2--1) emission-line ratios are consistent with star formation dominating the CO excitation over this range of transitions. The targets in our study represent a gas-rich phase of galaxy evolution with simultaneously high levels of star formation and nuclear activity; furthermore, the jets and outflows do not have an immediate appreciable impact on the global molecular gas reservoirs.
\end{abstract}




\section{Introduction}

The energy from accreting supermassive black holes (i.e. active galactic nuclei: AGN) is widely accepted to be responsible for the global quenching of star formation in massive galaxies \citep[AGN feedback; e.g. see reviews in][]{Alexander12,Fabian12,Harrison17}. However, the physical mechanisms by which this energy couples to the gas on galactic scales and its precise impact on the host galaxy remains unclear. Multi-wavelength studies are proving to be vital in both determining the mechanism and impact of feedback \citep[see e.g.][]{Cicone18,Cresci18}.

AGN are thought to be able to remove gas from their host galaxies via outflows. These outflows can be powered by the interaction between interstellar gas and small-scale accretion disc winds \citep{Faucher-Giguere12,Zubovas12} or directly via radiation pressure on dust \citep{Ishibashi15,Thompson15,Bieri17,Costa18,Costa18b}, particularly for AGN with high Eddington ratios (`quasar' or `radiative mode').
While typically thought to operate primarily by preventing hot halo gas from cooling, via the so called `radio' or `maintenance mode' \citep[e.g.][]{Churazov05}, collimated jets are also likely to drive outflows of interstellar gas \citep{Wagner12,Mukherjee16}, blurring the division between `quasar' and `maintenance' modes \citep[see e.g.][]{Jarvis19}.

In particular, the potential impact of AGN is most commonly observed through high velocity ionized gas outflows \citep[see e.g.][]{Karouzos16,Morganti17,Davies20}. However, if the direct impact of AGN upon star formation is to be understood, it is the cold ($\sim$10~K) molecular gas (primarily composed of H$_2$) which forms the fuel for star formation, that must be considered \citep{Morganti17}. Since cold molecular gas is not directly observable in H$_2$ emission, carbon monoxide ($^{12}$CO which has a permanent dipole moment), is most often used as a tracer of these cold molecular clouds \citep[see e.g.][and references therein]{Bolatto13,Carilli13}. Specifically, the ground level transition (J=1--0) has an excitation temperature of just 5.53~K, making it a good tracer of the total cold molecular gas \citep[see e.g.][]{Bolatto13}, while higher-J CO lines (i.e. J$\gtrsim$4--3) are produced from warmer, denser gas \citep[see e.g.][]{vanderWerf10,Daddi15,Mashian15,Kamenetzky17}.

 Molecular gas outflows traced by CO gas have been identified in both radio and quasar mode AGN \citep[see e.g.][]{Cicone14,King&Pounds15,Morganti15,Bischetti19,Fotopoulou19,Oosterloo19,Lutz20,Veilleux20}. However these outflows typically only represent $\sim$10~per cent of the molecular gas luminosity \citep[see e.g.][]{Fluetsch19,Lutz20} and so are difficult to observe. Instead, the impact of AGN on the molecular gas in their host galaxies is often probed through the total molecular gas content \citep[see e.g.][]{Bertram07,Xia12,Husemann17,Rosario18}. Specifically, the gas mass and the molecular gas fraction relative to the star formation rate are used to assess the potential impact of the AGN on the star formation efficiency and / or their ability to deplete the molecular gas supply within the host galaxies \citep[e.g.][]{Kakkad17,Perna18}. In addition to removing molecular gas through outflows, AGN and mechanical feedback from jets, can heat the molecular gas, which both inhibits star formation and causes the CO to emit in higher transitions \citep[see e.g.][]{Papadopoulos10}. 

Our recent results, combining integral field spectrographic (IFS) and radio observations, have identified a sample of luminous (L$_{\rm [O~III]}>$10$^{42}$\,erg\,s$^{-1}$) type 2 (obscured) AGN with signatures of jets and extended ionized gas outflows. These systems represent the ideal environment to search for signatures of feedback since they have the strong potential to interact with their environments both mechanically and radiatively \citep[][]{Harrison14, Harrison15,Lansbury18,Jarvis19}. In this work we use unresolved CO measurements of the (2--1) and (6--5) transitions, to investigate the molecular gas content of these systems and look for signatures of the impact of the AGN and jet in particular, through thermal excitation and depletion of the gas reservoir.  

In Section \ref{sec:sample} we introduce our sample, describe our spectral energy distribution (SED) fitting approach used to determine key galaxy and AGN properties (Section \ref{sec:SED_fitting}), and compare our sample to the star forming main sequence (Section \ref{sec:mainsequence}). Section \ref{sec:observations} describes our data, data reduction (Section \ref{sec:APEX_obs}) and analysis (Section \ref{sec:APEX_analysis}). In Section \ref{sec:results} we describe our results and in Section \ref{sec:discussion} we compare the total molecular gas and CO excitation in our systems to literature results (Sections \ref{sec:gas_content} and \ref{sec:excitation} respectively) and we discuss these results in the wider context of galaxy evolution in Section \ref{sec:galaxy_evolution}. We present our conclusions in Section \ref{sec:conclusions}. We adopt $H_0$=70km s$^{-1}$Mpc$^{-1}$, $\Omega_M$=0.3, $\Omega_\Lambda$=0.7 throughout, and assume a \citet{Chabrier03} initial mass function (IMF).

\section{Sample selection and properties}
\label{sec:sample}

Here we present observations of the molecular gas in nine type 2 quasars. This sample was designed to be representative of powerful local AGN, with signatures of feedback, and therefore is ideal for identifying the impact of the AGN on the molecular gas reservoir. In particular, by selecting sources with previously identified ionized gas outflows and radio jets, the AGN should be able to impact the molecular gas through radiative and / or mechanical feedback by exciting and / or removing the molecular gas. 

\begin{figure*}
 \centering
 \includegraphics[width=18cm]{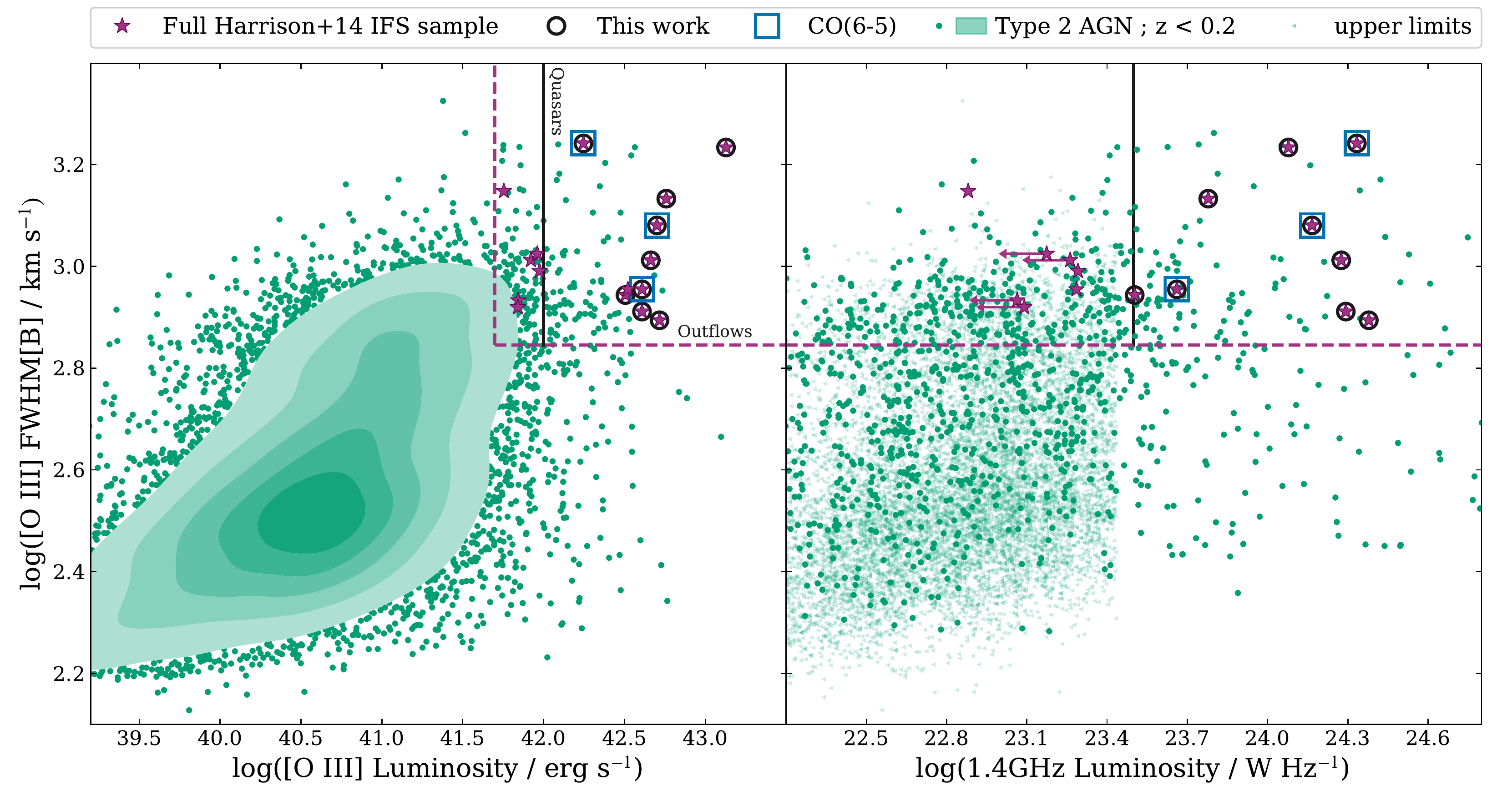}
 \caption{ This figure shows the basic sample properties and selection criteria. In each panel the full \citet{Harrison14} sample are shown as magenta stars with the sources studied here highlighted with black circles and the APEX CO(6--5) sample additionally highlighted with blue squares \citep[values tabulated in][]{Jarvis19}. Our parent population of $z<0.2$ type 2 AGN are shown as green data points and contours \citep{Mullaney13}. The dashed magenta lines show the selection criteria used in \citet{Harrison14} and the black solid lines mark the additional selection criteria applied for the sample in this work. \emph{Left}: The FWHM of the broadest, luminous \oiii\ emission-line component versus the total \oiii\ luminosity \citep[see][]{Harrison14,Jarvis19}. \emph{Right}: The FWHM versus the radio luminosity (from FIRST fluxes), where the parent sample sources with only upper limits on their radio luminosity are shown as light green triangles.
  }
  \label{fig:selection}
 \end{figure*}

In Fig.~\ref{fig:selection} we show how our targets were selected from the parent sample of 24\,264 $z<0.4$ spectroscopically identified AGN\footnote{using a combination of `BPT' diagnostics \citep{Baldwin81}, and emission-line widths.} presented in \citet{Mullaney13}. In \citet{Harrison14} we selected 16 $z<0.2$ type~2 AGN with luminous \oiii\ outflows: L$_{\rm [O~III]}>$10$^{41.7}$\,erg\,s$^{-1}$ and full width half maximum (FWHM) $\gtrsim$700~km~s$^{-1}$ (see Fig.~\ref{fig:selection}). The only other selection criteria was an ra / dec cut to select sources observable from Gemini-South. Using IFS data we revealed that these outflows are extended on $\gtrsim$kpc scales. For this work, we selected the nine of these 16 targets with the highest [O~{\sc iii}] luminosities (i.e.\ L$_{\rm [O~III]}>$10$^{42}$\,erg\,s$^{-1}$) and radio luminosities ($\log$[L$_{\rm 1.4GHz}$/W\,Hz$^{-1}$]$\geq$23.5; see Fig.~\ref{fig:selection}). In \citet{Jarvis19} we established that the radio luminosity of these targets is dominated by emission from the AGN, with eight of the nine exhibiting extended radio structures on 1--25~kpc scales, which are likely radio jets (J1010$+$0612 is the only target without any evidence for an extended radio structure). The spatial coincidence of these radio features to outflows and disturbed ionised gas features visible in the IFS data strongly suggests jet--gas interactions in the majority of this sample \citep[see][]{Harrison15,Jarvis19}. The basic sample properties are provided in Table~\ref{tab:targets}.

\subsection{SED fitting}
\label{sec:SED_fitting}

A significant amount of the analysis in this paper relies on having reliable estimates of the star formation rates and stellar masses in our systems without contamination from the AGN. Since the AGN in this work are all type 2, the AGN has only a small contribution to the \emph{UV} -- optical emission but may still contribute significantly to the infrared emission. As such, in \citet{Jarvis19} we performed \emph{UV}--\emph{IR} SED fitting using the `Code Investigating GALaxy Emission' \citep[CIGALE\footnote{\url{https://cigale.lam.fr}};][]{Noll09,Buat15,Ciesla15} to derive the host galaxy and AGN properties of this sample. We used data from GALEX, SDSS, 2MASS, WISE, IRAS, and where available archival, {\it Herschel} PACS and SPIRE, for these SED fits, corrected for Galactic extinction \citep[see][]{Jarvis19}. Specifically, CIGALE simultaneously fits the attenuated stellar emission, star formation heated dust emission, AGN emission (from the accretion disc and dust heating) and nebular emission. Of particular relevance for this work are the stellar mass (M$_\star$) and the star formation rate (SFR) of the host galaxies, which are listed in Table~\ref{tab:targets}. We calculated these star formation rates from the SED-derived infrared luminosity due to star formation ($L_{\rm IR,SF}$) and the relationship from \citet{Kennicutt12}, converting from a Kroupa to a Chabrier IMF by multiplying by 0.94 \citep{Madau14}, specifically: SFR=\lir/(2.57$\times10^{43})\times$0.94, with \lir\ in erg s$^{-1}$ and SFR in M$_\odot$~yr$^{-1}$. For further details of the SED fitting, derived quantities and uncertainties we refer the reader to \citet{Jarvis19}. 

In Table~\ref{tab:targets} the quoted uncertainties are 1$\sigma$ formal errors from the CIGALE fits and do not include systematics. However, there is a 0.3~dex systematic uncertainty expected on the infrared luminosity and stellar mass from the SED fitting \citep{Gruppioni08,Mancini11,Santini15}. This results in a 0.42~dex systematic uncertainty for the star formation rate values, from adding in quadrature the systematic uncertainties from the SED fitting and the 0.3~dex systematic uncertainty on the conversion between \lir\ and SFR \citep{Kennicutt12}. Our sources have stellar masses in the range 8$\times 10^9<$~M~$_\star$$<$~$1.1$$\times 10^{11}$M$_\odot$ and star formation rates in the range 8~$<$~SFR~$<$~84~M$_\odot$yr$^{-1}$. All but one of our sources (J1430$+$1339) are classified as Luminous Infrared Galaxies (LIRGs) based upon their infrared luminosities due to star formation (10$^{11}$--10$^{12}$~L$_{\odot}$; see Fig.~\ref{fig:MS}).

We verified these values by performing independent SED fits using another code: AGNfitter \citep{CalistroRivera16}. The main difference between this code and the CIGALE code is that CIGALE assumes an energy balance between the \emph{IR} and optical emission for the host galaxy, where AGNfitter considers the two almost independently with a prior that the energy from the \emph{IR} must be at least equal to the energy attenuated from the stellar emission. For the three of our sources with {\it Herschel} observations, which have the best available coverage of the \emph{FIR} emission \citep[namely J1100$+$0846, J1356$+$1026 and J1430$+$1339; see][]{Jarvis19}, the \emph{IR} derived SFRs agree within 0.27~dex (i.e. within the systematic uncertainty). For the remaining six sources the best SFR from AGNfitter is based on the optical emission alone and as such is a lower limit on the actual SFR \citep{CalistroRivera16}. In each case this limit is consistent with our SFR from CIGALE. The stellar masses from AGNfitter are on average 0.19~dex higher than those from CIGALE (i.e. within the systematic uncertainty), and the only significant outliers are J1000$+$1242 and J1010$+$0612 which have AGNfitter derived stellar masses 0.67 and 0.48~dex larger than those from CIGALE respectively. We note that using the AGNfitter stellar masses and SFRs would not change the main conclusion of this work. In summary, we trust the CIGALE SED-derived stellar masses and SFRs used throughout this work, within the limitations of the unavoidable systematic uncertainties discussed above.  

\subsection{Our targets in the context of the star-forming main sequence}\label{sec:mainsequence}

\begin{figure}
 \centering
 \includegraphics[width=\hsize]{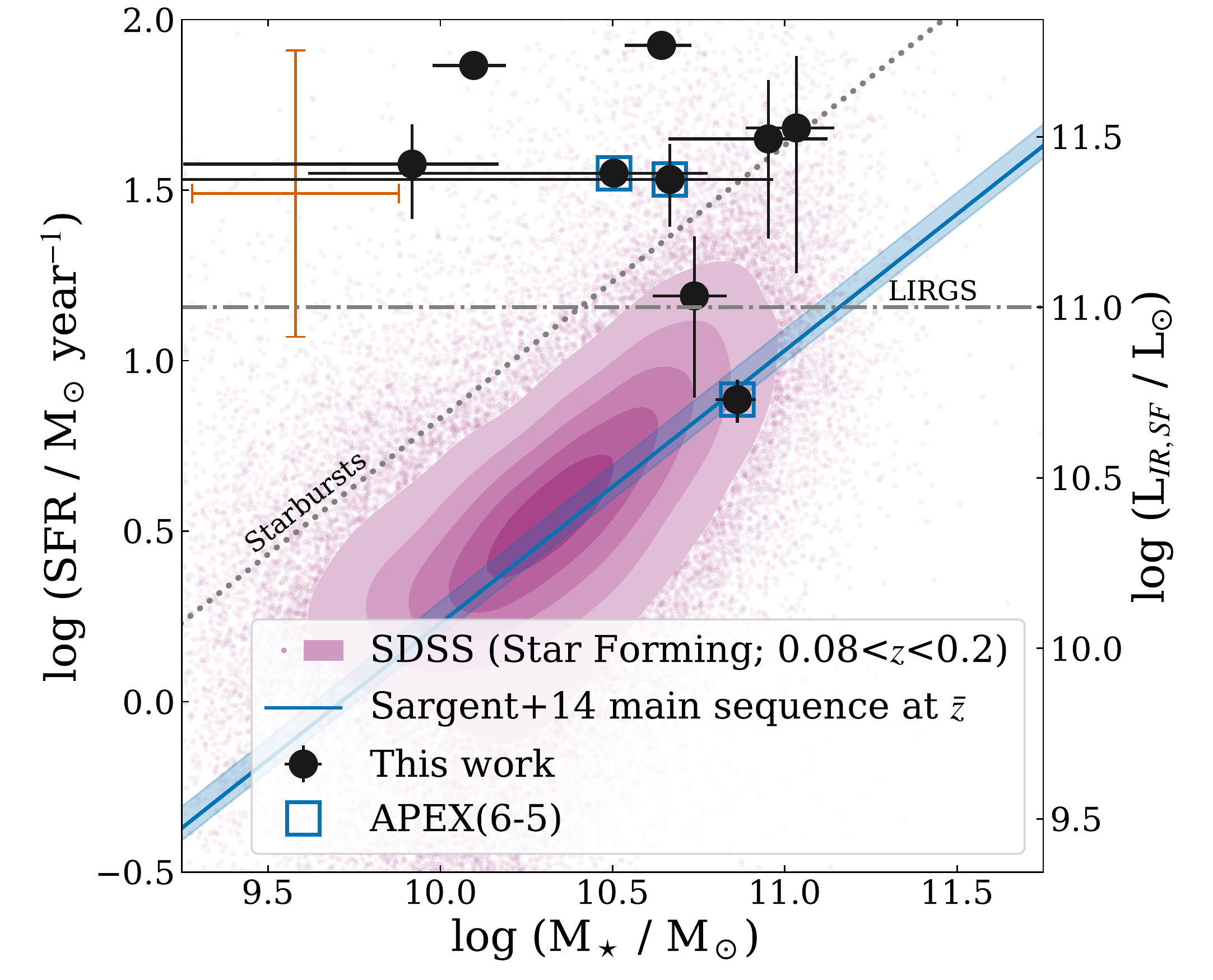}
 \caption[]{ Our sources compared to the star forming main sequence, as shown by the comparison of stellar mass (M$_\star$) and star formation rate (SFR). Our sample is shown as black circles and the APEX CO(6--5) sample is highlighted with blue squares. The red error bar in the top left corner shows the systematic errors (see Section~\ref{sec:SED_fitting}). The magenta contours and small translucent points show values for star-forming galaxies from SDSS. The solid blue line is the main sequence as given in \citet{Sargent14} at the mean redshift of our sources ($z$=0.127), with the width showing the variation across the redshift spanned by our sources. The black dotted line marks the region occupied by starbursts ($\Delta_{\textrm{MS}}$>4) and the black dot-dashed lines marks the limit for LIRGs (L$_\textrm{IR}$$\geq$$10^{11}$~L$_\odot$). All of our sources lie on or above the main sequence with seven being classified as starbursts.}
  \label{fig:MS}
 \end{figure}

 There is a long established trend observed between star formation rate and stellar mass for star-forming galaxies, which is commonly referred to as the ``star-forming main sequence'' \citep[see e.g.][]{Brinchmann04,Daddi07,Elbaz07,Noeske07,Salim07,Wyder07}. This relation provides a useful comparison to put our sample into the wider context of star-forming galaxies. Specifically, we consider where our galaxies lie in comparison to the redshift dependent main sequence of \citet{Sargent14} (see Fig.~\ref{fig:MS}). We chose this parametrization as it visually provided the best fit to galaxies selected from SDSS (the parent sample of our work); where the other main sequences checked were: \citet{Bauermeister13,Speagle14,Whitaker14,Genzel15,Schreiber15}. In Fig.~\ref{fig:MS} we show the \citet{Sargent14} main sequence compared to all SDSS sources defined as star-forming based on BPT emission-line ratios \citep{Kauffmann03b,Brinchmann04,Tremonti04} within $z=0.08$--0.2 (i.e.\ the redshift range spanned by our sources) using the MPA-JHU measurements\footnote{\url{https://www.sdss.org/dr12/spectro/galaxy_mpajhu/}} converted from a Kroupa to Chabrier IMF \citep{Madau14}. We define the distance from the main sequence ($\Delta_{\textrm{MS}}$) for each source as the ratio of its specific star formation rate (sSFR $\equiv$ SFR/M$_\star$) compared to that of the main sequence at its redshift and stellar mass \citep{Sargent14}. Following the literature, we define our targets as ``starbursts'' if they have $\Delta_{\textrm{MS}}$>4 \citep[see e.g.][]{Elbaz11}; however, we note that we use this definition for a comparison to the overall population only and do not claim that they are physically different to the rest of the population for this work.

Using the definitions described above, all of our sources are on or above their local main sequence with seven classified as starbursts\footnote{J1010$+$1413 is right on the transition between normal star forming and starburst with \deltaMS=3.7. In the rest of this paper we assume \deltaMS$\sim$4 and consider it as a starburst.}, even though we applied no pre-selection on SFR or infrared luminosity (see Table \ref{tab:targets}).

\begin{table*}
 \caption{Target list and properties.
 \newline Notes: (1) Object name; (2)--(3) optical RA and DEC positions from SDSS (DR7); (4) Systemic redshifts from GMOS IFS data \citep[\oiii;][]{Jarvis19}; (5)--(8) are directly from, or are derived from, the CIGALE SED fits first presented in \citet{Jarvis19} and discussed here in Section \ref{sec:SED_fitting}: (5) Stellar mass from SED fitting (there is an additional $\sim$0.3~dex systemic uncertainty not included in the quoted errors); (6) Infrared luminosity from star formation in the range 8--1000 $\mu$m (i.e.\ excluding the AGN contribution; there is a $\sim$0.3~dex systemic uncertainty not included in the quoted errors); (7) Star formation rate calculated from L$_{\text{IR,SF}}$ (there is a $\sim$0.42~dex systematic uncertainty not included in the quoted errors; see Section \ref{sec:SED_fitting}); (8) distance of the source from the \citet{Sargent14} main sequence, defined as sSFR/sSFR$_{MS}$ (see Section~\ref{sec:mainsequence}); (9) Ratio of the \oiii 5007 to H$\beta$ emission lines from SDSS DR7 catalogues \citep[single Gaussian fits;][]{Abazajian09}. Additional details of these sources (e.g. radio luminosity and AGN bolometric luminosity) are given in \citet{Jarvis19}. 
 }

	\begin{tabular}{c c c c c c c c c } 
	\hline
   
	 Name & RA & Dec  & $z$ & log(M$_{\star}$) & $\log(L_{\text{IR,SF}})$ & SFR & \deltaMS  & \oiii/H$\beta$ \\ 
   & (J2000) & (J2000) & & (M$_\odot$) & (erg s$^{-1}$)  & (M$_\odot$ yr$^{-1}$) & &  \\
 (1) & (2) & (3) & (4) & (5)& (6) & (7)& (8) & (9)  \\
	\hline   
J0945+1737 & 09:45:21.33 & +17:37:53.2 & 0.1281 & 10.1$^{+0.09}_{-0.12}$ & 45.3$\pm$0.02 & 73$\pm$4 & 36.1 & 1.015$\pm$0.005 \\
J0958+1439 & 09:58:16.88 & +14:39:23.7 & 0.1091 & 10.74$^{+0.09}_{-0.12}$ & 44.6$^{+0.2}_{-0.3}$ & 15$\pm$8 & 2.4 & 1.124$\pm$0.005 \\
J1000+1242 & 10:00:13.14 & +12:42:26.2 & 0.1479 & 9.9$^{+0.3}_{-0.7}$ & 45.0$^{+0.1}_{-0.2}$ & 40$\pm$10 & 24.8 & 0.988$\pm$0.008 \\
J1010+1413 & 10:10:22.95 & +14:13:00.9 & 0.1992 & 11.0$\pm$0.1 & 45.1$^{+0.2}_{-0.4}$ & 50$\pm$30 & 3.7 & 1.1$\pm$0.005 \\
J1010+0612 & 10:10:43.36 & +06:12:01.4 & 0.0982 & 10.5$^{+0.3}_{-0.9}$ & 44.99$\pm$0.04 & 35$\pm$3 & 8.7 & 0.828$\pm$0.005 \\
J1100+0846 & 11:00:12.38 & +08:46:16.3 & 0.1004 & 10.7$^{+0.3}_{-2.4}$ & 45.0$\pm$0.1 & 34$\pm$9 & 6.2 & 1.098$\pm$0.005 \\
J1316+1753 & 13:16:42.90 & +17:53:32.5 & 0.1504 & 11.0$^{+0.2}_{-0.3}$ & 45.1$^{+0.2}_{-0.3}$ & 40$\pm$20 & 4.3 & 1.082$\pm$0.005 \\
J1356+1026 & 13:56:46.10 & +10:26:09.0 & 0.1233 & 10.64$^{+0.09}_{-0.11}$ & 45.36$\pm$0.02 & 84$\pm$4 & 15.3 & 0.982$\pm$0.004 \\
J1430+1339 & 14:30:29.88 & +13:39:12.0 & 0.0852 & 10.86$^{+0.05}_{-0.06}$ & 44.32$^{+0.06}_{-0.07}$ & 8$\pm$1 & 1.0 & 0.883$\pm$0.004 \\

	\hline   
	\end{tabular}
    
\label{tab:targets} 

	\end{table*}

\section{Observations and data reduction}
\label{sec:observations}

\subsection{Data reduction}

\label{sec:APEX_obs}

We use the Atacama Pathfinder EXperiment (APEX) to observe the spatially un-resolved molecular gas emission in the CO(2--1) and CO(6--5) transitions. We observed CO(2--1) for the whole sample presented here and CO(6--5) for a representative sub-sample (J1010$+$0612, J1100$+$0846 and J1430$+$1339; see Figs.~\ref{fig:selection} and \ref{fig:MS}). These specific transitions were selected based on a combination of scientific and observational constraints. Specifically, lower CO transitions (CO(1--0) in particular) are best used to trace the total cold molecular gas content \citep[e.g.][]{Bolatto13,Carilli13} and the CO(2--1) transition is the lowest observable at the redshift of our targets with the available APEX instrumentation. Higher CO transitions trace molecular gas that has been excited by star formation, shocks and the AGN \citep[e.g.][]{Mashian15,Carniani19,Vallini19}. Specifically the CO(6--5) transition was selected based on indications that it can be boosted by AGN activity and jets in particular \citep{Papadopoulos10}, and because it was the highest transition that could be observed for our targets in a reasonable time using APEX, due to available instrumentation and the atmospheric transmission. Due to observing constraints (e.g. the need for good weather for these observations; see Table \ref{tab:apex_data}) we only observed three of our targets in CO(6--5), however these three are representative of the overall population (see Figs.~\ref{fig:selection} and \ref{fig:MS}).   

We observed CO(2--1) for our targets under proposal id.\ E-0100.B-0166 [PI: Jarvis] with the observations carried out between 2017 July 7 and 2018 December 29 with precipitable water vapours (PWV) between 0.6 and 4.7~mm. Three different instruments were used for these observations due to the redshift range of the targets and changes in the available instrumentation over the period of observation, namely, the Swedish-ESO PI receiver for APEX \citep[SEPIA180;][]{Belitsky18}, the Max Planck Institute for Radio Astronomy's PI230 and the APEX-1 receiver \citep[SHeFI 230~GHz band;][]{Vassilev08}. The instrument used for each source, the dates they were observed and the PWV values during the observations are listed in Table \ref{tab:apex_data}. The CO(6--5) data was observed under proposal id.\ E-0104.B-0292 [PI: Harrison] and observed between 2019 August 31 and December 10 using the SEPIA660 band 9 instrument with PWVs between 0.4 and 0.6.

The data were reduced and analysed using the Continuum and Line Analysis Single-dish Software ({\sc class}; version mar19a).\footnote{from the {\sc GILDAS} software package \url{http://www.iram.fr/IRAMFR/GILDAS/}} For many of our sources spectral spikes (due to bad channels) were found in at least one polarisation. To correct for this while losing the minimum amount of data for each source, for each day of observations we examined the average spectrum from each of the spectrometers separately and flagged, by eye, any channels affected by spikes. We also flagged the leading 150 channels (80 for the CO(6--5) data) and trailing 10 (in the overlap region) of each individual spectra. We combined the two spectrometers in the same sideband and polarisation using Zhiyu Zhang's {\sc class} extension file: combineTwoIFsAPEX.class which is made available online at \url{https://github.com/ZhiyuZhang/gildas_class_libraries}. From each of these combined scans we subtracted a linear baseline using the {\sc class} {\sc base} command, excluding a velocity range $\sim$500~km~s$^{-1}$ to either side of the observed line position or the expected line position from the SDSS redshift if no line was obviously seen in the total binned spectrum. We then removed scans with poor baselines based on the ratio of their rms in 50~km~s$^{-1}$ bins (selected to best reveal the baselines) compared to the theoretical rms (rms$_{t}$) calculated by the following equation: \begin{equation} \textrm{rms}_{t} \equiv \frac{T_{sys}}{\sqrt{|d\nu\times 10^6\times t|} }, \label{eq:rrms} \end{equation} where $T_{sys}$ is the system temperature, $t$ is the integration time and $d\nu$ is the frequency step size. The cutoff value for each was selected based on a combination of visual examination and minimizing the resultant final rms of the combined data in 100~km~s$^{-1}$ bins (selected to best reveal the emission lines) and ranged from rms / rms$_{t}$=1.25--2. Each day's data were then multiplied by the appropriate Kelvin to Jansky conversion factor. For each time frame and instrument the K/Jy conversion was determined using the APEX telescope efficiencies tool (\url{http://www.apex-telescope.org/telescope/efficiency/}), supplemented by private communications with Juan-Pablo Perez-Beaupuits (see Table \ref{tab:apex_data} for the values used). Finally, the spectra were combined into a single spectrum and re-sampled to 1~km~s$^{-1}$ bins with a final linear baseline removed. 

We show the final reduced APEX data in the velocity range around the CO(2--1) emission line in Fig.~\ref{fig:APEX_data} and around the CO(6--5) emission line in Fig.~\ref{fig:APEX_6-5_data}.

\begin{table}
 \caption{Details of the observations
 \newline Notes: This table is divided into two parts with the details of our APEX CO(2--1) data given first and our CO(6--5) data at the bottom. (1) Object name; (2) Instrument; (3) On source time of the final total spectrum; (4) Date observed (year-month-day); (5) Conversion factor used to convert the observed antenna temperature (in K) to flux density (in Jy); (6) Average precipitable water vapour (PWV; mm) during the observations.}

	\begin{tabular}{c c c c c c} 
	\hline
   
	 Target & Instrument & t$_\textrm{on}$  & Date & K/Jy & pwv \\ 
	 & & (min) &  & & (mm) \\
 (1) & (2) & (3) & (4) & (5) & (6) \\
	\hline   
    \multicolumn{6}{l}{CO(2--1); proposal id. E-0100.B-0166 [PI: Jarvis]}\\
    \hline 
J0945$+$1737 & SEPIA180 & 126 & 2018-10-24 & 36$\pm$5 & 0.7\\
& & & 2018-11-11 & 36$\pm$5 & 1.4\\
& & & 2018-12-27 & 36$\pm$5 & 2.1\\
& & & 2018-12-28 & 36$\pm$5 & 4.7 \\
& & &  2018-12-29 & 36$\pm$5 & 3.0\\
J0958$+$1439 & SEPIA180 & 71.6 & 2018-10-26 & 36$\pm$5 & 0.6\\
& & & 2018-12-27 & 36$\pm$5 & 2.1 \\
& & & 2018-12-28 & 36$\pm$5 & 4.7\\
J1000$+$1242 & SEPIA180 &  261 & 2018-10-24  & 36$\pm$5 & 0.7 \\
& & & 2018-11-01  & 36$\pm$5 & 1.3\\
& & & 2018-11-02 & 36$\pm$5  & 1.3 \\
& & & 2018-11-04  & 36$\pm$5  & 1.5\\
& & & 2018-11-05 & 36$\pm$5 & 0.7\\
& & & 2018-11-08 & 36$\pm$5 & 0.9 \\
J1010$+$1413 & SEPIA180 & 28.5 & 2018-10-26  & 36$\pm$5 & 0.6\\
& & & 2018-10-31 & 36$\pm$5 & 0.9 \\
& & & 2018-11-02 & 36$\pm$5 & 1.3 \\
& & &  2018-11-03  & 36$\pm$5  & 1.5\\
J1010$+$0612 & PI230  & 167& 2018-10-29 & 42$\pm$6 & 0.8 \\
J1100$+$0846 & SEPIA180 & 51.4  & 2017-07-27 & 40$\pm$6 & 0.9 \\
& & &  2018-11-03  & 36$\pm$5 & 1.5\\
J1316$+$1753 & SEPIA180 & 78.5 & 2017-07-28 & 40$\pm$6 & 0.9\\
& & & 2018-12-28  & 36$\pm$5 & 4.7\\
J1356$+$1026 & SEPIA180 & 35.8  & 2017-07-27 & 40$\pm$6 & 0.9 \\
& & & 2017-07-28 & 40$\pm$6 & 0.8 \\
J1430$+$1339 & APEX-1  & 101 & 2017-07-29 & 38$\pm$6  & 0.7\\
& & & 2017-07-30 & 38$\pm$6 & 0.9\\
& & & 2017-07-31  & 38$\pm$6 & 1.0 \\ 
& & & 2017-08-02  & 38$\pm$6 & 0.7\\
& & & 2017-08-03   & 38$\pm$6 & 0.9\\
& & & 2017-08-31  & 38$\pm$6 &1.7\\
& & & 2017-09-01  & 38$\pm$6 & 1.6 \\
	\hline   
    \multicolumn{6}{l}{CO(6--5); proposal id. E-0104.B-0292 [PI: Harrison]}\\
    \hline   
J1010$+$0612 & SEPIA660 & 145.2 & 2019-10-29 & 69$\pm$6 & 0.5 \\ 
J1100$+$0846 & SEPIA660 & 220.6 & 2019-11-05 & 69$\pm$6 & 0.4 \\
& & & 2019-11-06 & 69$\pm$6& 0.4 \\
& & & 2019-12-10 & 69$\pm$6& 0.5 \\
J1430$+$1339 & SEPIA660 & 118 & 2019-08-31 & 69$\pm$6& 0.6\\
    
	\hline   
	\end{tabular}
    
\label{tab:apex_data} 

	\end{table}

\subsection{Data analysis}
\label{sec:APEX_analysis} 
We fit each averaged spectrum using Bayesian fitting and MCMC implemented through {\sc emcee} \citep{Foreman-Mackey13}.\footnote{\url{http://dfm.io/emcee/current/}} This Bayesian method is preferred over frequentist fitting techniques for this analysis since it is less sensitive to binning, provides realistic uncertainties, and for the upper limits in particular requires only minimal assumptions on the line profile (see Appendix \ref{app:bayesian}). We assume a single Gaussian profile for the line, and fit for the line flux (f; integral under the line), peak velocity ($v_p$; central line velocity offset from the systemic redshift in Table \ref{tab:targets}), and standard deviation ($\sigma$; the width of the line) as well as the standard deviation of the noise in the spectrum ($\sigma_N$) which we assume to be Gaussian. The results of the fitting are listed in Table \ref{tab:apex_fits} and shown in Fig.~\ref{fig:APEX_data} and \ref{fig:APEX_6-5_data}. The full details of this analysis are given in Appendix \ref{app:bayesian}.

 \begin{figure*}
 \centering
 \includegraphics[width=17cm]{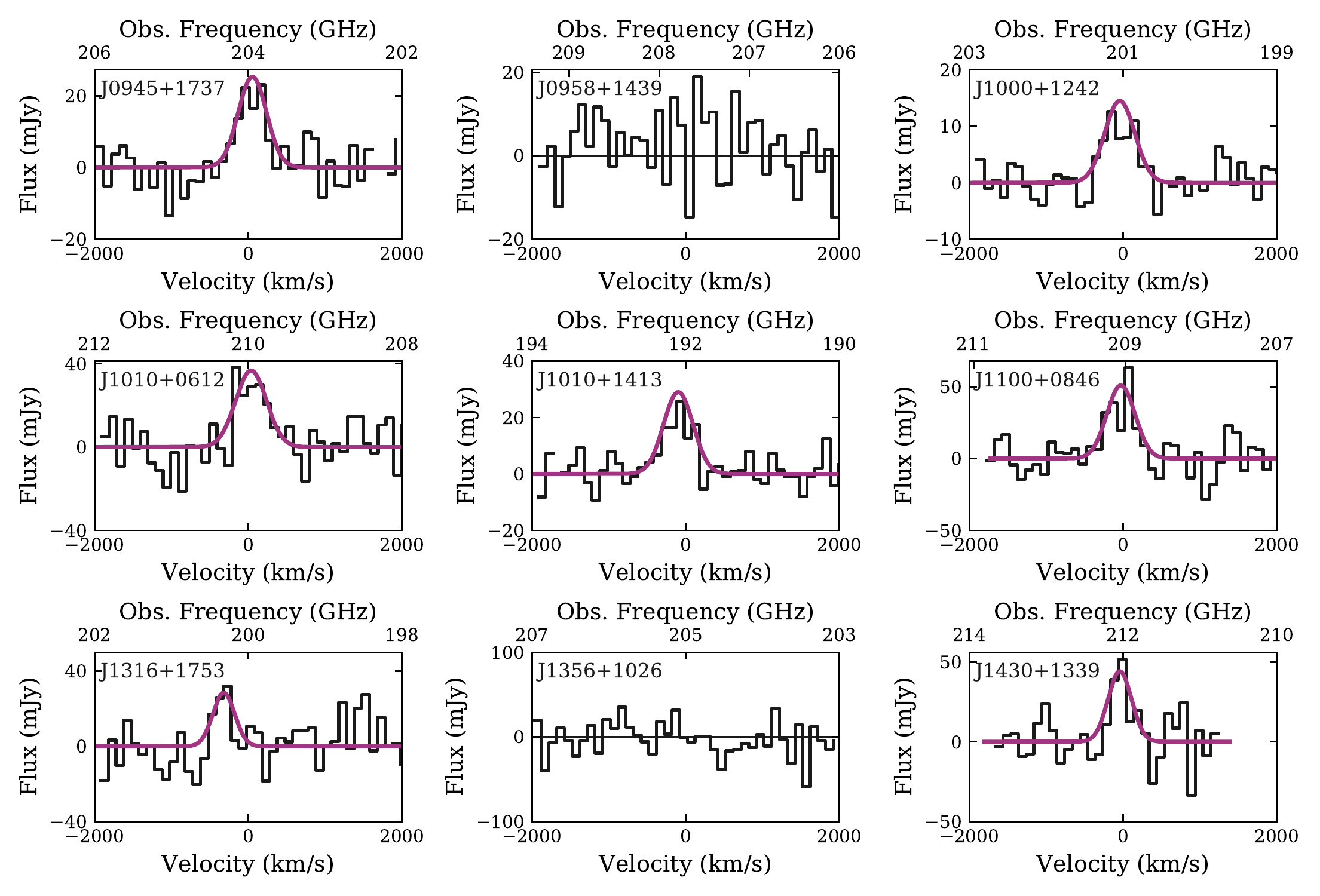}
 \caption{ Our APEX CO(2--1) data (black curve) for each source. Over plotted for each is the results of our Bayesian fitting to the emission line, specifically the Gaussian constructed from the 50th percentile value from the posteriors for each parameter (magenta; see Table~\ref{tab:apex_fits}). For the two non detections a black horizontal line at flux=0 is plotted to help guide the eye. 
  }
  \label{fig:APEX_data}
 \end{figure*}
 \begin{figure*}
 \centering
 \includegraphics[width=17cm]{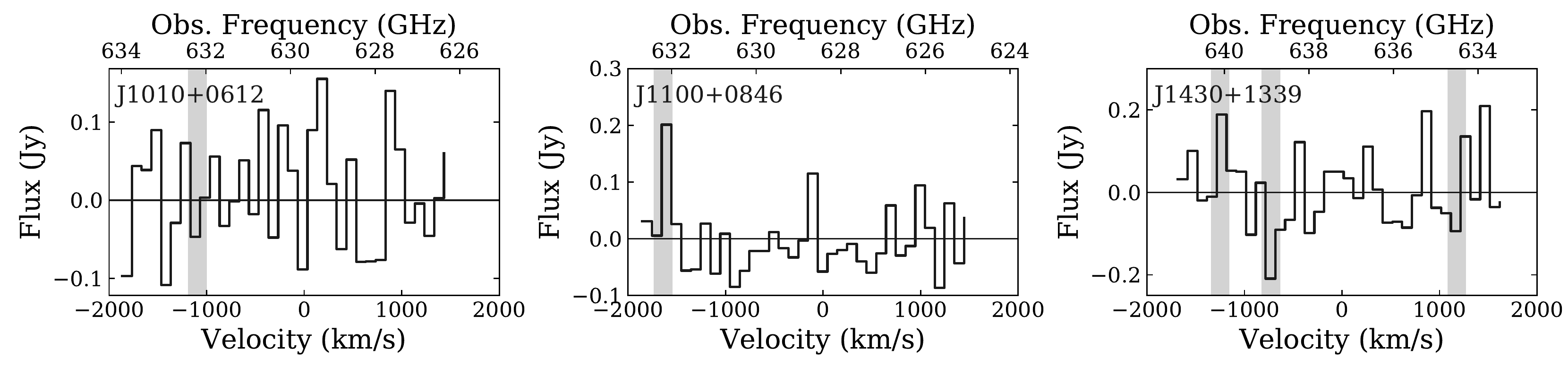}
 \caption{Our APEX CO(6--5) data. All three are undetected. Grey vertical bands highlight frequencies where there are narrow atmospheric absorption features that can cause slightly higher noise. A black horizontal line at flux=0 is plotted to help guide the eye. 
  }
  \label{fig:APEX_6-5_data}
 \end{figure*}

\subsection{Evaluating contamination from other sources and beam corrections}
\label{sec:ALMA_check}

The beams of the APEX observations discussed here are $\sim$28~arcsec ($\sim$52~kpc at a representative redshift of $z=0.1$) for the CO(2--1) data and $\sim$9~arcsec ($\sim$17~kpc at $z$=0.1) for the CO(6--5) data. Based on the relatively large beams of the APEX data and considering the optical sizes of our targets we do not expect any CO flux to fall beyond our observed beams, making beam corrections unnecessary. However, the large CO(2--1) beam raises the possibility that other CO bright objects may be contaminating our flux measurements. To check for this scenario, we used higher spatial resolution ALMA CO observations. Specifically, we use the CO(1--0) and CO(3--2) images published in \citet{Sun14} for J1356$+$1026 and for the other targets we use preliminary CO(3--2) images from two proposals carried out by our group\footnote{Specifically, id.\ 2016.1.01535.S (PI. Lansbury), and id.\ 2018.1.01767.S (PI. Thomson)} which have spatial resolution of $\sim$0.3--0.5~arcsec and a maximum recoverable scale of $\sim$4~arcsec.\footnote{For J1430$+$1339, the ALMA data has a maximum recoverable scale of $\sim$19~arcsec.} The only target where a possible contaminating CO source was identified is J1010$+$0612 which has a CO(3--2) bright companion $\sim$7~arcsec away, which is within our CO(2--1) beam. Preliminary flux measurements from the ALMA data reveal that $\sim$82~per cent of the total flux is in our primary target of J1010$+$0612, a difference which is within the 1$\sigma$ error bars from our Bayesian fit.\footnote{This is supported by the percent of the total system flux in J1010$+$0612 from 2MASS which is 87, 92 and 91~per cent, of the combined fluxes of these two sources, in the J, H and K$_S$ bands, respectively.} We highlight this source in subsequent figures.

\section{Results}
\label{sec:results}

We show the final reduced APEX data, in 100~km~s$^{-1}$ bins, around the CO(2--1) line (for all nine targets) in Fig.~\ref{fig:APEX_data} and around the CO(6--5) line in Fig.~\ref{fig:APEX_6-5_data} (for the three targets observed). In the online supplementary data for this paper we provide corner plots displaying the posterior probability distributions of each of the parameters for each source.
For the CO(2--1) data all but J0958$+$1439 and J1356$+$1026 show distinct peaks in the probability distribution for each parameter, indicating a detection. Therefore we detect seven of our nine targets in CO(2--1). None of the three sources observed in CO(6--5) show distinct peaks in the posterior probability distributions of all parameters and are clearly undetected. 

For the detected emission lines we quote the 50th percentile (median) of the posterior distribution for each parameter in Table~\ref{tab:apex_fits}, and use the 16th and 84th percentile as errors. We note there is an additional $\sim$13~percent systematic uncertainty on the line flux from the error on the temperature to flux density conversion factors (Section~\ref{sec:APEX_obs}). The values derived from our Bayesian analysis are consistent within errors to those derived from fitting a Gaussian directly to the data in 100~km~s$^{-1}$ bins. In Fig.~\ref{fig:APEX_data} we show the resulting line profiles from our Bayesian procedure as Gaussians constructed using the 50th percentile value for each parameter. These parameter values will be adopted for the analyses throughout this work. 

For the non-detected emission lines we derived 3$\sigma$ upper limits on the line flux from the 99.7th percentile on the posterior distribution (see values in Table~\ref{tab:apex_fits}). Our upper limit for J1356$+$1026 (i.e.\ \lco$_{(2-1)}$<6$\times$10$^9$~K~\kms~pc$^2$) is consistent with the observed value obtained by converting the total \lco(1--0) reported for this source in \citet{Sun14} to CO(2--1) (i.e.\ \lco(2--1)$=$0.82$\times$10$^9$~K~\kms~pc$^2$), where we have assumed  \lco(2--1)/(1--0)$\equiv$r$_{21}$=0.8 (see Section \ref{sec:derived_quantities} for a discussion of the choice of r$_{21}$). J1356$+$1026 is discussed in more detail in Section \ref{sec:gas_content} and Section \ref{sec:galaxy_evolution}. We have no prior knowledge of the total CO emission for J0958$+$1439. 

Overall we detect the CO(2--1) line for seven of our nine targets with fluxes in the range 7--23~Jy~\kms. The two non detected targets have upper limits of 21.5 and 33.1~Jy~\kms\ (for J0958$+$1439 and J1356$+$1026, respectively). Our upper limits on the CO(6--5) fluxes are 110, 74 and 135~Jy~\kms\ for J1010$+$0612, J1100$+$0846 and J1430$+$1339, respectively. For the CO(2--1) detections we measured peak line velocities between $-$320 and 50~\kms\ relative to the systematic redshifts in Table \ref{tab:targets}, and line widths ($\sigma$) between 150 and 200~\kms; however, we defer a discussion of the molecular gas kinematics to future work. 

\begin{figure}
 \centering
 \includegraphics[width=\hsize]{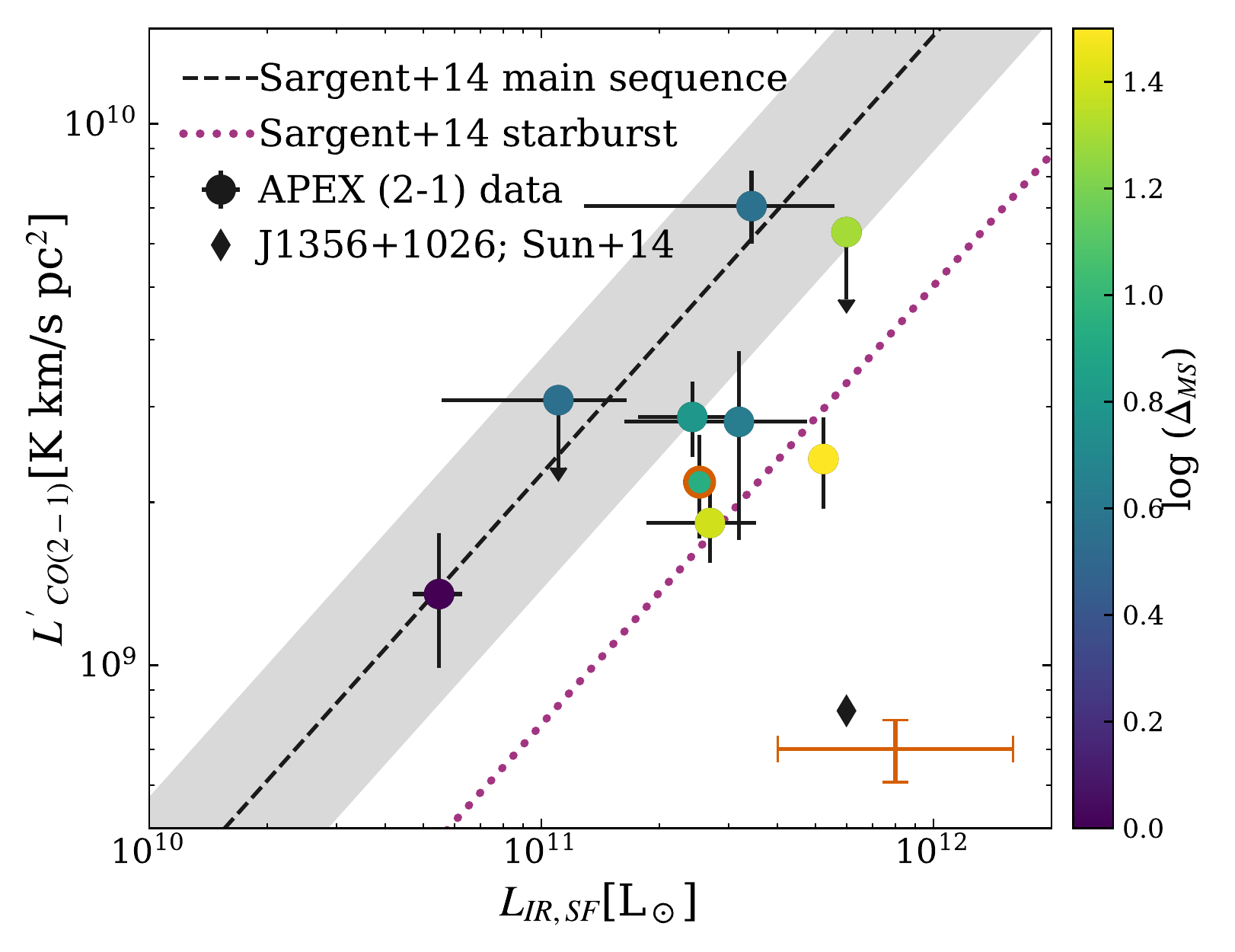}
 \caption{ \lco(2--1) compared to the infrared emission produced by dust-heated star formation between 8--1000~$\mu$m (L$_\textrm{IR,SF}$; see Section~\ref{sec:SED_fitting}). Our sample are shown as circles, colour coded by their distance from the main sequence (\deltaMS; see section \ref{sec:mainsequence}). In the bottom right is a representative error bar showing the systematic uncertainties (see Sections \ref{sec:SED_fitting} and \ref{sec:results}). The \citet{Sun14} value for \lco(1--0), converted to (2--1) using r$_{21}$=0.8, for J1356$+$1026 is shown as a black diamond. The black dashed and magenta dotted lines show the relationships from \citet{Sargent14} for main sequence and starburst galaxies, respectively (see Section~\ref{sec:lco_vs_lir}). Our quasars appear to follow the trend of star-forming galaxies, with those further from the main sequence agreeing more closely with starburst relation. J1010$+$0612 is highlighted with a red outline because the \lco\ may be $\lesssim$18~per cent overestimated (see Section \ref{sec:APEX_analysis}).   
  }
  \label{fig:L'CO_vs_LIR}
 \end{figure}

We calculate the CO luminosities \citep[following e.g.][]{Solomon97} for each source using: 
\begin{equation}
 \textrm{L}^\prime_\textrm{CO} [\textrm{K km s}^{-1}\textrm{pc}^2]=     \frac{3.25\times10^7}{\nu_\textrm{co,rest}^2}\left(\frac{D_L^2}{1+z}\right)\textrm{f},
\end{equation}
where $D_L$ is the luminosity distance in Mpc, $\nu_\textrm{co,rest}$ is the rest-frame frequency of the CO line in GHz (230.538 and 691.473~GHz for the CO(2--1) and CO(6--5) lines respectively), and f is the flux of the CO line in Jy~\kms. This results in \lco(2--1) values of (1.4--7)$\times 10 ^9$~K km s$^{-1}$ pc$^2$ for the seven detected sources (see Table \ref{tab:apex_fits}). These are plotted as a function of infrared luminosity in Fig.~\ref{fig:L'CO_vs_LIR} and are discussed in the following section.

\begin{table*}
 \caption{CO emission-line measurements
 \newline Notes: This table is divided into two parts with the details of our fits to the APEX CO(2--1) data given first then our fits to the CO(6--5) data are in the bottom portion. (1) Object name; (2--5) are values derived from our Bayesian fits to the APEX data, consisting of the 50th percentile (median) value with errors derived from the 16th and 84th percentiles: (2) line flux in Jy. For non-detections 3$\sigma$ upper limits are given; (3) Peak velocity in km~s$^{-1}$ with respect to the systematic redshift given in Table \ref{tab:targets}; (4) Width of the line as a standard deviation in km~s$^{-1}$; (5) Standard deviation of the noise in the final 1~km~s$^{-1}$ binned spectrum (see Section~\ref{sec:APEX_analysis}); (6) \lco / 10 $^9$ in K km s$^{-1}$ pc$^2$.
 \newline $^{\dagger}$ Due to a nearby CO bright companion which is included within the CO(2--1) beam, the true CO(2--1) flux of this source could be up to 18~per cent lower than the value given here (see Section~\ref{sec:ALMA_check}; the other line parameters are not used in the discussion of this paper). 
 }

	\centering  
	\begin{tabular}{l c c c c c } 
	\hline
   
	\multicolumn{1}{c}{ Name} & f & $v_p$ & $\sigma$ & $\sigma_\textrm{N}$ & \lco \\ 
   & (Jy~km~s$^{-1}$) & (km~s$^{-1}$) & (km~s$^{-1}$) & (Jy) &   (1$\times 10 ^9 \times $ K km s$^{-1}$ pc$^2$)\\
\multicolumn{1}{c}{ (1)} & (2) & (3) & (4) & (5) & (6)\\
	\hline   
	    \multicolumn{6}{l}{CO(2--1)}\\
    \hline 
 
J0945+1737 & 12$\pm$2 & 50$^{+50}_{-40}$ & 180$^{+50}_{-40}$  &  0.062 & 2.3$^{+0.5}_{-0.4}$\\
J0958+1439 & $<$21.5 & -- & -- & 0.086 &  $<$3.0\\
J1000+1242 & 7$\pm$1 & -40$\pm$40 & 200$^{+40}_{-30}$  & 0.031 &  1.9$\pm$0.3 \\
J1010+1413 & 14$\pm$2 & -100$\pm$30 & 190$^{+40}_{-30}$   &  0.11  &  7$\pm$1\\
J1010+0612$^{\dagger}$ & 19$\pm$4 & 30$^{+60}_{-50}$ & 200$^{+60}_{-50}$   &  0.052  &  2.1$^{+0.5}_{-0.4}$\\
J1100+0846 & 23$\pm$4 & -30$\pm$40 & 180$^{+40}_{-30}$    & 0.12   &  2.8$\pm$0.4\\
J1316+1753 & 10$^{+5}_{-4}$ & -320$^{+90}_{-60}$ & 140$^{+110}_{-50}$  &  0.13   & 3$\pm$1\\
J1356+1026 & $<$33.1 & -- & --    & 0.23    &  $<$6.0 \\
J1430+1339 & 17$\pm$5 & -40$^{+60}_{-50}$ & 150$^{+60}_{-40}$   &  0.16    &1.4$\pm$0.4 \\

	\hline   
	    \multicolumn{6}{l}{CO(6--5)}\\
    \hline 
J1010$+$0612 & $<$110 & -- & --   & 0.56  & $<$1.4  \\
J1100$+$0846 & $<$74 & -- & --   & 0.48   &   $<$0.98 \\
J1430$+$1339 & $<$135 & -- & --  &  0.90  &   $<$1.2\\
	\hline   
	\end{tabular}
    
\label{tab:apex_fits} 

	\end{table*}

\section{Discussion}
\label{sec:discussion}

In this work we look for signatures of AGN feedback on the molecular gas in our quasar sample. They are luminous AGN with ionized outflows and jets which may be able to impact upon the gas supply either radiatively or mechanically \citep[see e.g.][for a review]{Harrison17}. We stress that although molecular outflows are commonly observed directly through broad, generally blue shifted emission line components \citep[see e.g.][]{Fluetsch19,Lutz20}; they are typically weak in CO emission (contributing $\lesssim$10~per cent of the total emission-line profile), which would be undetectable in our data. Here we focus on the galaxy-wide molecular gas content (Section \ref{sec:gas_content}) and CO excitation (Section \ref{sec:excitation}) of our sample of extreme quasars and compare them to redshift-matched literature galaxy samples both with and without AGN.

\subsection{Molecular gas content}
\label{sec:gas_content}

In order to assess if our AGN have depleted their host galaxies' gas reservoir or decreased their star formation efficiency, we compare our results to studies of general galaxy populations and other AGN samples. Specifically, we consider: (1) their total CO luminosities (\lco) compared to their infrared luminosities (\lir; corrected for the AGN contribution; Section \ref{sec:lco_vs_lir}); (2) how the molecular gas fractions (\Mgas/M$_\star$) and depletion times (\Mgas/SFR) compare to other samples when star-formation rates, stellar masses and offsets from the star-forming main sequence (\deltaMS) are taken into account (section \ref{sec:derived_quantities}) and (3) the relationship between AGN properties and the molecular gas content and star formation of the host galaxy (Section \ref{sec:AGN}).

\subsubsection{\lco\ / \lir\ relations }
\label{sec:lco_vs_lir}

The correlation of \lco\ (which traces the molecular gas mass) and \lir\ (which traces star formation) in star-forming galaxies is well studied (e.g. \citealt{Kennicutt98,Genzel10,Greve14,Sargent14}). By directly comparing observable quantities, this analysis removes many of the assumptions that are needed to convert these values into physical parameters. A complication to this analysis, which is not always accounted for, is that AGN can contribute significantly to the \emph{IR} emission \citep[see e.g.][]{Kirkpatrick19}. The careful SED fitting technique implemented in our work allows us to reliably consider only the IR luminosity from the star formation component which is free from AGN contamination (i.e.\ $L_{\rm IR,SF}$; see Section \ref{sec:SED_fitting}).

Numerous works have parametrized the \lco\ -- \lir\ relation using different samples of galaxies and different CO transitions. Here we focus on the work of \citet{Sargent14} which used CO observations of 130 $z$$<$$3$ massive (M$_\star>10^{10}$~M$_\odot$) star-forming and starburst galaxies collected from a range of surveys. The size of the \citet{Sargent14} sample and its coverage of similar galaxy properties as in this work make it an ideal comparison sample. They find a redshift-invariant log-linear relation between the \lco\ and \lir. We convert their relation from CO(1--0) to CO(2--1) using r$_{21}$=0.8 (\citealt{Leroy09,Sargent14,Daddi15,Tacconi18})\footnote{\citet{Sargent14} use r$_{21}$=0.8 to convert from observed CO(2--1) to CO(1--0) in their analysis (where needed). Possible biases introduced by the choice of r$_{21}$ are discussed in Section \ref{sec:derived_quantities}.}. We compare the \lco\ and \lir\ values for the nine targets in our sample to this relation in Fig.~\ref{fig:L'CO_vs_LIR}.

We find that two out of the seven CO(2--1) detected quasars are consistent with the \lco\ -- \lir\ relationship for main sequence star-forming galaxies, whilst the other five have \lco\ values up to a factor of $\sim$4 lower than the relation would predict for their \lir\ (see Fig.~\ref{fig:L'CO_vs_LIR}). However, as highlighted by the colour scaling in Fig.~\ref{fig:L'CO_vs_LIR}, all of the targets with low \lco\ compared to the \citet{Sargent14} main sequence relationship have high star formation rates in relation to the main sequence (i.e.\ they have high \deltaMS\ values; see Section \ref{sec:mainsequence}). This is consistent with \citet{Sargent14}, which finds that starbursts are offset to lower \lco values by a factor of $\sim$2.9, on average, compared to main sequence galaxies (see dotted line in Fig.~\ref{fig:L'CO_vs_LIR}). Indeed, all of our quasars which fall below the \lco -- \lir\ relationship for main sequence galaxies are classified as starbursts (i.e.\ \deltaMS$\gtrsim4$) and fall within 0.3\,dex of the \citet{Sargent14} relationship for starburst galaxies. We note that similar results are found when comparing our sample to the LIRG and merger \lco\ -- \lir\ relationships of \citet{Greve14} and \citet{Genzel10}, respectively.

Based on our data, the two CO(2--1) non-detected targets could still be consistent with the expected relationships for star-forming galaxies (the main sequence and starburst relations for J0958$+$1439 and J1356$+$1026, respectively); but could also lie significantly lower. Specifically, we note that using the \citet{Sun14} CO(1--0) luminosity for J1356$+$1026 would place it $\sim$4 times lower than the \citet{Sargent14} starburst relation (see Fig.~\ref{fig:L'CO_vs_LIR}). This source is discussed in more detail in Section \ref{sec:derived_quantities} and Section \ref{sec:galaxy_evolution}. 

In summary, we find that at least seven of our nine targets have \lco\ values consistent with those of the star-forming galaxy population at matched infrared luminosities and at similar distance to the main sequence. From this analyses there is no evidence that the observed ionized outflows and jets in our powerful quasars have had an instantaneous impact on the observed CO luminosities.

\subsubsection{Molecular gas comparisons}
\label{sec:derived_quantities}

The more physically motivated quantities to study are the gas fraction (ratio of the molecular gas mass to stellar mass) and the depletion time (ratio of the molecular gas mass to the star formation rate), which relates to how efficiently stars are being formed for a given molecular gas mass. Based on large galaxy samples, these molecular gas properties scale with redshift, stellar mass and distance from the star-forming galaxy main sequence \citep[see e.g.][and references therein]{Tacconi18,Liu19}. In this work we are not concerned with the physical significance of these relations, but use them as a tool to compare the molecular gas properties of our sample to the wider galaxy population. 
 
\begin{figure*}
 \centering
 \includegraphics[width=18cm]{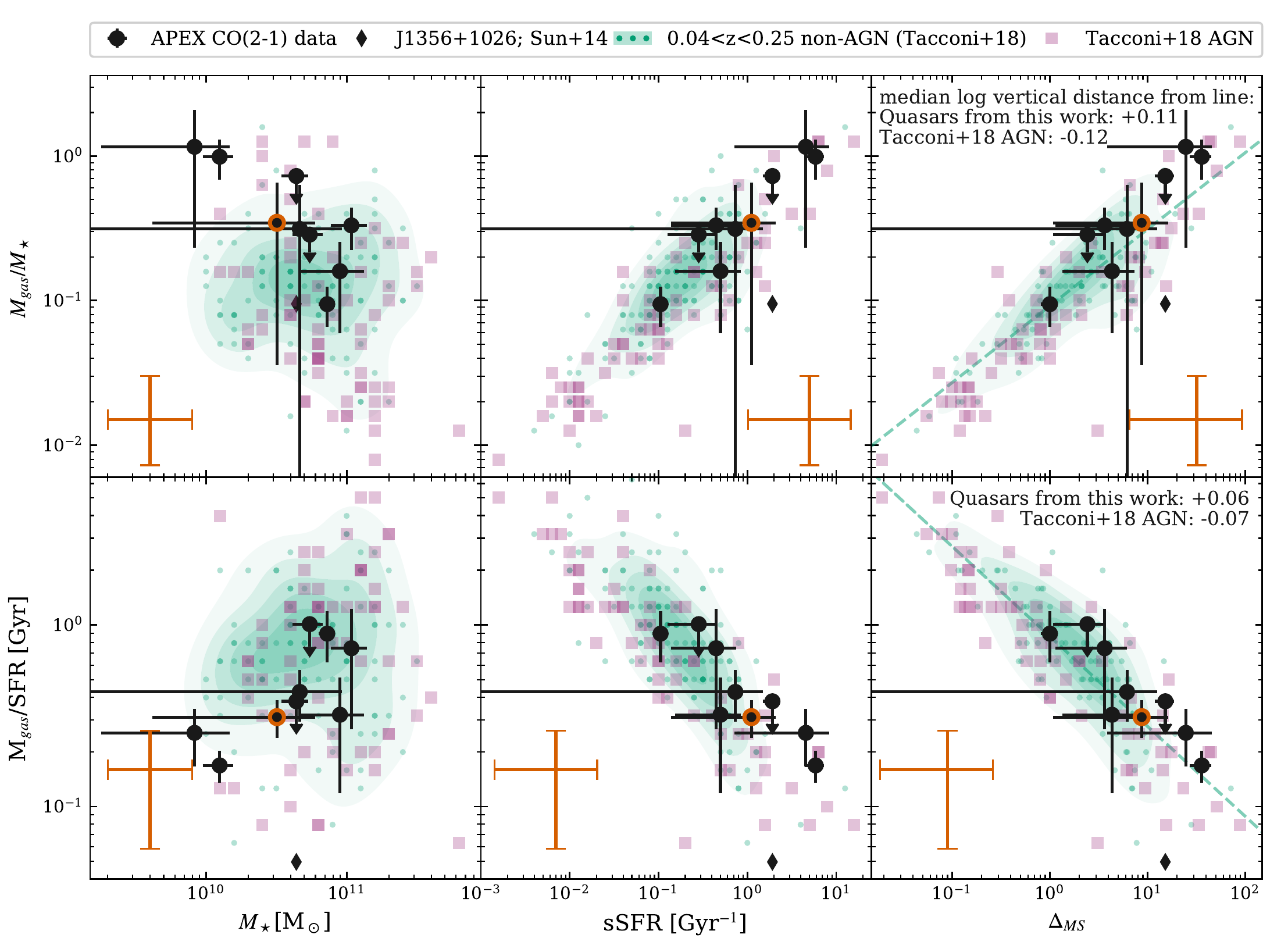}
 \caption{ A comparison of the molecular gas content of our quasars (black circles) to literature values taken from \citet{Tacconi18}, within $z$~$\pm$0.05 of the full range of redshifts spanned by our sample. Galaxies without an identified AGN are represented by green points and density contours and AGN host galaxies by magenta squares (see Section \ref{sec:derived_quantities}). We show how the molecular gas fractions (\Mgas/M$_\star$; top row) and depletion times (\Mgas/SFR; bottom) vary with: stellar mass (M$_\star$; left column), sSFR (middle column) and distance to the main sequence (\deltaMS; right column; see Sect.~\ref{sec:sample}). J1010$+$0612 is highlighted with a red circle following Fig.~\ref{fig:L'CO_vs_LIR} and a black diamond in each panel marks the value for J1356$+$1026 using \lco(1--0) from \citet{Sun14} instead of the limit from this work. In each panel a representative error bar is shown which factors in the systematic errors that could cause relative shifts between this work and the comparison sample (i.e.\ the conversion from \lir\ to SFR and the error on the K/Jy conversion from APEX; see Sections~\ref{sec:SED_fitting} and \ref{sec:results}). In the \deltaMS\ column (right) the median log vertical distance from a linear fit to the \citet{Tacconi18} non-AGN (green dashed line) is given. Our powerful CO detected quasars, containing both outflows and jets, follow the overall trends seen in the comparison sample in all panels.
  }
  \label{fig:derived_qualtities}
 \end{figure*}

We compare our data to the homogenised sample of \citet{Tacconi18} limited to within $\pm$0.05 of the maximum and minimum redshift of our sample and only using their CO based measurements.\footnote{Combined CO(1--0) and (3--2).} Specifically, the data compiled comes from the xCOLD GASS \citep{Saintonge17}, EGNOG \citep{Bauermeister13} and GOALS \citep{Armus09} surveys and from the sample presented in \citet{Combes11}. We identified AGN hosts for each sample using BPT-based AGN classifications (the same as used to identify our sample; see Section~\ref{sec:sample}), where available, and including all AGN classes \citep[e.g.\ LINERS, Seyferts, quasars, composite;][]{Baldwin81}. The galaxies in this redshift-matched comparison sample span the complete range of stellar mass, sSFR and \deltaMS\ found for our sample (see Fig.~\ref{fig:derived_qualtities}).

To ensure consistency with the comparison sample, we calculate the molecular gas masses of our samples using the same procedure as in \citet{Tacconi18}. Specifically, we follow the metallicity dependent \alphaco\ and mass-metallicity relation used by \citet{Tacconi18} \citep[see also][]{Genzel15} to calculate molecular gas masses following \Mgas= \alphaco$\times$\lco(1--0). The resultant \alphaco\ values for our sample range from 4.0 to 4.2. We convert from \lco(2--1) to \lco(1--0) using r$_{21}$=0.8. The full details of the equations used and a table of derived values are presented in Appendix \ref{app:alphaca}. For our seven CO(2--1) detected targets, the derived molecular gas masses fall in the range of 9.9$<$log(\Mgas/M$_\odot$)$<$10.5, with corresponding ranges of gas fractions and depletion times of \Mgas/$M_{\star}$=0.1--1.2 and \Mgas/SFR$=$0.16-0.95~Gyr, respectively.

In Fig.~\ref{fig:derived_qualtities} we compare our derived gas masses and depletion times to the \citet{Tacconi18} population as a function of stellar mass, sSFR and \deltaMS. We note that the dependence on the choice of main sequence relation adds additional uncertainty to \deltaMS\ compared to sSFR; however, \deltaMS\ has been shown to be more closely related to the molecular gas properties \citep[see e.g.][]{Tacconi18,Liu19} and we obtain consistent conclusions if we just consider sSFR. Within errors, our sources overlap with the comparison sample (non-AGN and AGN) in all of the common diagnostic planes shown in Fig.~\ref{fig:derived_qualtities}. To quantify this comparison, we perform a simple log linear fit to the \citet{Tacconi18} sample with AGN removed (see Fig.~\ref{fig:derived_qualtities}). Our sample have a median log vertical offset of $+$0.1 in the gas fraction versus \deltaMS\ plane and $+$0.04 in the depletion time versus \deltaMS\ plane (ignoring the non-detections).\footnote{Where the log vertical offset for a point (a,b) from a line y=f(x) with both in log space, is defined as b-f(a).} This provides some evidence for moderately high ($\sim$0.1\,dex) gas fractions in our sample, with respect to their position relative to the main sequence. However, we can not rule out that the two non detected sources in our sample could bring our average down. Specifically, calculating the gas mass for J1356$+$1026 using the total \lco(1--0) from \citet{Sun14} would place it among the most gas poor systems in the \citet{Tacconi18} population, with a log vertical distance from the \citet{Tacconi18} line of $-$0.61 and $-$0.66 in gas fraction and depletion time respectively (see Section \ref{sec:galaxy_evolution}).

The AGN included in \citet{Tacconi18}, which have no selection for high bolometric luminosity or outflows, go in the opposite direction to our CO-detected targets, with median log vertical offsets of $-$0.12 in the gas fraction versus \deltaMS\ plane and $-$0.07 in the depletion time versus \deltaMS\ plane. We explore the possible role of AGN power further in Section~\ref{sec:AGN}.

It is important to consider possible systematic uncertainties in comparing AGN to non AGN samples due to the assumptions required to calculate gas masses.\footnote{We also note that, although the \citet{Tacconi18} work does not directly account for an AGN contribution to their stellar mass and SFR calculations, their sample does not include type 1 quasars and are typically low power AGN (and therefore the AGN do not dominate the optical--UV part of the SEDs) so the impact is not expected to be strong.} For example, there is no consensus on if AGN have systematically different ratios of \lco(2--1) and (1--0), which is used to convert between the two (r$_{21}$; see e.g.\ \citealt{OcanaFlaquer10,Papadopoulos12,Xia12,Husemann17,Shangguan19}); however, we note that the observed range is modest (0.4$<$r$_{21}$$<$$1.2$) and we have adopted the mean value of 0.8 throughout this work \citep[see e.g.][]{Braine93,Leroy09}. A larger uncertainty comes from \alphaco, which, for most galaxies appears to have a value of $\sim$4, with slight dependencies on metallicity and SFR \citep[see e.g.][and references therein]{Bolatto13,Sandstrom13}. However \alphaco\ may be significantly lower in LIRGs, submillimetre galaxies, mergers, starbursts and AGN \citep[as low as $\sim$0.6; see e.g.][]{Bolatto13,Sargent14,Calistro-Rivera18}. In our comparison to literature results we have controlled for many of these differences, i.e.\ we are comparing like-for-like in sSFR and \deltaMS\ and made consistent assumptions (see Appendix \ref{app:alphaca}). However, we can not rule out some level of systematic differences in \alphaco\ for AGN which could shift our sources to systematically lower gas masses than the non-AGN comparison sample. Finally, we note that a limitation of our comparison to the \citet{Tacconi18} catalogue is that it does not provide information on detection fractions or report upper limits. However, if anything, this limitation will strengthen our suggestion that the majority of the quasars in our sample, are comparatively gas rich.

To summarize, although we can not control for unknown systematic variations in \alphaco, our quasar sample has molecular gas fractions and depletion times that are consistent with, or slightly higher than, the redshift matched comparison sample when considered in terms of their stellar masses, sSFRs or distances to the main sequence. This implies no significant rapid depletion of the molecular gas supply despite the presence of kpc ionized gas outflows and jets.

 \subsubsection{The impact of AGN on the molecular gas content}
 \label{sec:AGN}
 
\begin{figure}
 \centering
 \includegraphics[width=\hsize]{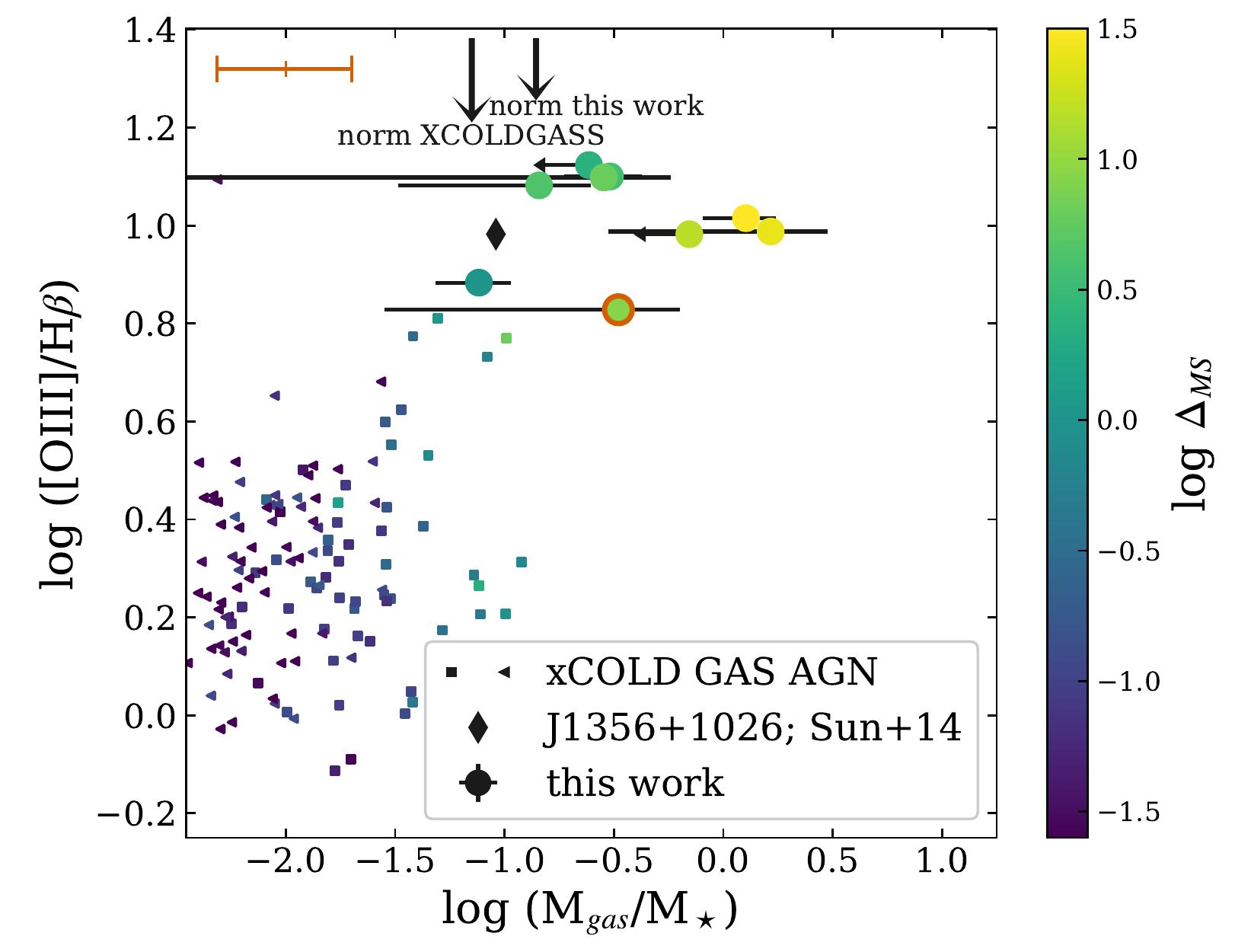}
 \caption{\oiii~5007 to H$\beta$ emission-line ratio versus gas fraction (\Mgas/M$_\star$) of our quasars (circles) and xCOLD GASS AGN \citep[squares, with triangles for upper limits; from the catalogues provided with][]{Saintonge17}. The errors on the \oiii / H$\beta$ ratios for our sample are smaller than the point size. A representative systematic error bar is shown in the top left as in Fig.~\ref{fig:derived_qualtities}. J1010$+$0612 is highlighted with a red outline as per Fig.~\ref{fig:L'CO_vs_LIR} and a black diamond marks the value for J1356$+$1026 using \lco(1--0) from \citet{Sun14} instead of the limit from this work. Data points are colour-coded by their distance to the \citet{Sargent14} main sequence (\deltaMS). The two arrows show the average gas fraction from the \citet{Tacconi18} sample matched in stellar mass and redshift to each population, to demonstrate that the observed trend is not dominated by differences in these parameters between the two samples. Sources with high \oiii\ / H$\beta$ ratios tend to have high gas fractions and have sSFR's above the main sequence.
  }
  \label{fig:AGN_comp}
 \end{figure}

To investigate the relationship between AGN and the molecular gas content in more detail, we build upon the work of \citet{Saintonge17} which found that the BPT selected AGN in the xCOLD GASS sample with the highest \oiii\ / H$\beta$ ratios (taken as a proxy of the power of the AGN radiation field) tend towards higher gas fractions. In Fig.~\ref{fig:AGN_comp} we plot gas fractions as a function of the \oiii\ / H$\beta$ ratio for both the xCOLD GASS sample and the quasars presented in this work. For a fair comparison with the \citet{Saintonge17} data we, again, use r$_{21}$=0.8, and follow their method to obtain \alphaco. That is, we use the metallicity and \deltaMS\ dependent function of \citet{Accurso17}, which results in \alphaco\ values between 3.3 and 6.0, and gas masses of 9.8<log(\Mgas/M$_\odot$)<10.5 (for the seven detected targets; see Appendix \ref{app:alphaca} for full details).\footnote{We note that J0945+1738, J1000+1242 and J1356$+$1026, are strong starbursts and might be better described with a lower \alphaco\ (see Appendix \ref{app:alphaca}); however, this would not change our conclusions.}

Fig.~\ref{fig:AGN_comp} reveals that our sample, extending to the most extreme local AGN, with no pre-selection on molecular gas or star forming properties, agrees with and strengthens the previous results from xCOLD GASS: the more extreme AGN (i.e.\ those with log (\oiii\ / H$\beta$)$\gtrsim$0.6) tend to have the highest gas fractions. On average, for the combined samples we find (\Mgas/M$_\star)_\textrm{average}$=0.02 for the sources with log(\oiii / H$\beta$)$<$0.6 and (\Mgas/M$_\star)_\textrm{average}$=0.16 for the sources with log(\oiii/H$\beta$)$>$0.6 (excluding non-detections). We note that a similar trend is observed for our sample when the bolometric AGN luminosity from our SED fits \citep[see][]{Jarvis19} is used instead of \oiii\ / H$\beta$.  

 As highlighted by the colour-coding in Fig.~\ref{fig:AGN_comp}, the most extreme AGN with the highest gas fractions are hosted in galaxies with high levels of concurrent star formation. Specifically we find increasingly high \deltaMS\ values for increasing \oiii\ / H$\beta$ values. Indeed when considering instead of the gas fraction, the log vertical offset of each AGN from linear fits to redshift and stellar mass matched \citet{Tacconi18} samples in the gas fraction versus \deltaMS\ plane (see Section \ref{sec:derived_quantities}), the trend with \oiii / H$\beta$ disappears. Specifically, the median vertical offset of the combined sample with log(\oiii/H$\beta$)$<$0.6 is $-$0.05 while the median value for the sources with log(\oiii/H$\beta$)$>$0.6 is $\sim$0 (see also Section \ref{sec:derived_quantities} and Fig.~\ref{fig:derived_qualtities}). We note however, that our sample covers a very narrow range of \oiii\ / H$\beta$ and xCOLD GAS is not designed as an AGN survey and so due to volume and redshift limits does not contain any powerful AGN. Larger samples, uniformly covering AGN with a range of powers would be needed to strengthen this observation. We discuss the impact of these results on the relationships between AGN activity, molecular gas masses and star formation rates in Section~\ref{sec:galaxy_evolution}.

\subsection{CO excitation}
\label{sec:excitation}

\begin{figure}
 \centering
 \includegraphics[width=\hsize]{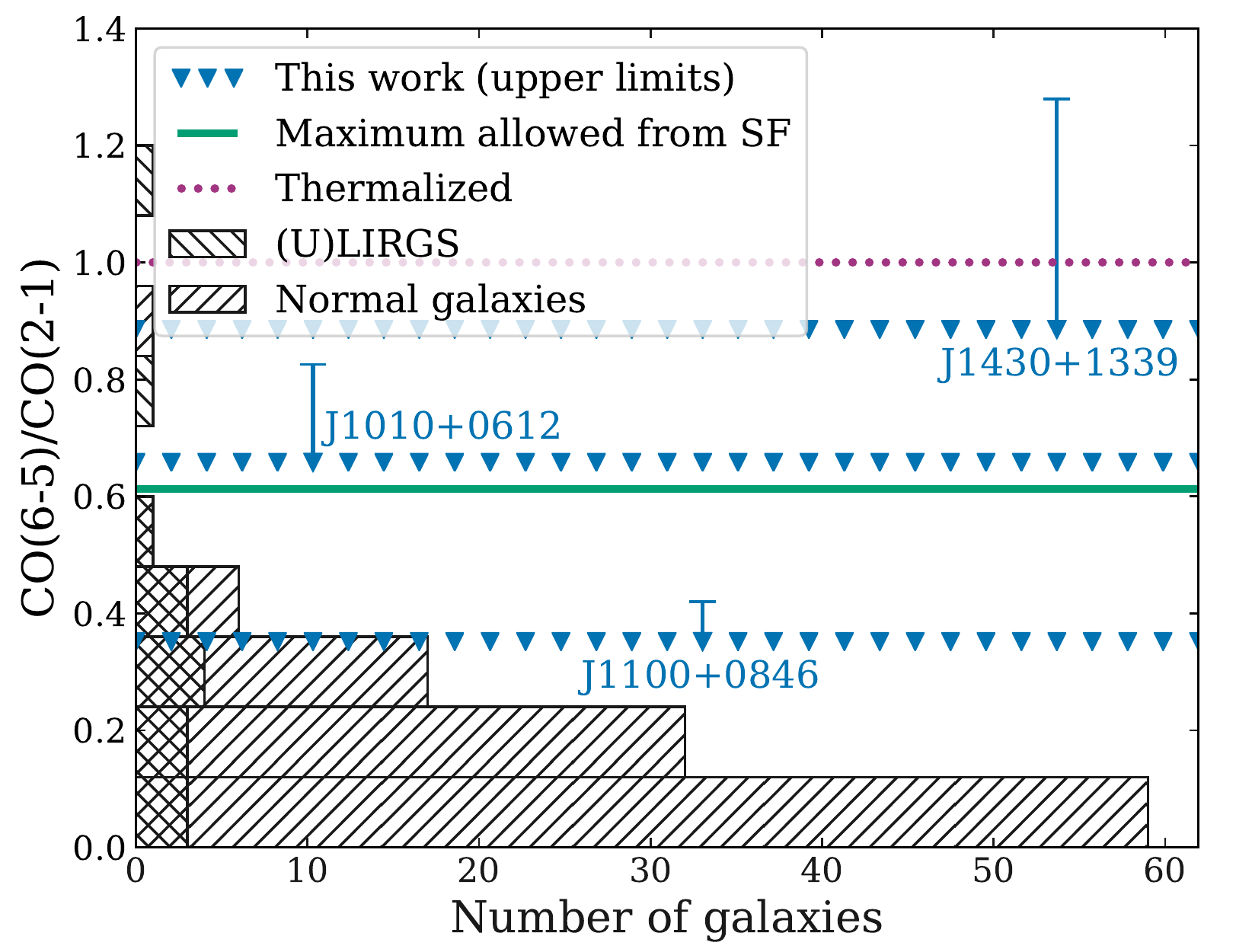}
 \caption{Upper limits of \lco (6--5) / \lco (2--1) ratios, for the three sources with these observations, represented as lines of blue triangles. Corresponding error bars represents the maximum values given the uncertainty on the measured CO(2--1) flux. The green line marks a maximum ratio achievable from star formation alone assuming a maximum star formation rate surface density of 1000~M$_\odot$yr$^{-1}$kpc$^{-2}$ following \citet{Narayanan14}. The dotted magenta line marks where the molecular gas becomes thermalized. The histograms show the distribution of \lco (6--5) / CO(2--1) ratios for literature galaxy samples from \citet{Papadopoulos12} ([U]LIRGs only) and \citet{Kamenetzky16}. For at least two of our sources (J1010$+$0612 and J1100$+$0846) we do not have any evidence for highly excited CO SLEDs (see Sec.~\ref{sec:excitation}).}
  \label{fig:excitation}
 \end{figure}

The relative luminosity of different CO lines contains information about the conditions of the molecular gas and the mechanisms that are exciting it. Through our APEX observations we put constraints on the ratio of the CO(6--5) to the CO(2--1) luminosity (\lco; r$_{62}$) for three sources in our sample. Specifically, we find r$_{62}$$<$$0.66$, 0.35 and 0.89 for J1010$+$0612, J1100$+$0846 and J1430$+$1339 respectively (see Fig.~\ref{fig:excitation}). For J1010$+$0612 if an 18~per cent lower CO(2--1) flux is assumed to account for possible blending with its close companion, the limit on r$_{62}$ increases marginally to 0.8 (which is within the error bar shown in Fig.~\ref{fig:excitation}).

The most ubiquitous source of CO excitation is photodissociation regions (PDRs) from the UV photons emitted from young stars. However this mechanism is inefficient at exciting higher CO transitions. Shocks and / or X-ray emission (through X-ray-Dominated Region models; XDR), both of which can be powered by AGN or jets, are needed to further excite the CO gas \citep[see e.g.][]{Pereira-Santaella13,Carniani19}.

The CO spectral line energy distribution (SLED) modelled by \citet{Narayanan14}, which depends solely on the star formation rate surface density ($\Sigma_\textrm{SFR}$), predict values of r$_{62}$$\lesssim$$0.24$ for typical star formation rate surface densities of $\lesssim$10~M$_\odot$yr$^{-1}$kpc$^{-2}$, and even for an exceptionally high limit of $\Sigma_\textrm{SFR}$=1000~M$_\odot$yr$^{-1}$kpc$^{-2}$, r$_{62}$ $\gtrsim$ 0.6 can not be achieved. Our observed limit for J1100$+$0846 in particular suggests that the excitation of its total molecular gas could be explained by star formation alone, even at the highest end of the possible r$_{62}$ ratio for this source. For J1010$+$0612 the observed limit of r$_{62}$$<$$0.66$ could be explained by star formation alone; however, some contribution of shocks and XDR, possibly powered by the AGN can not be ruled out. 

In Fig.~\ref{fig:excitation} we also show the distribution of observed r$_{62}$ ratios from \citet{Kamenetzky16} and \citet{Papadopoulos12}. This shows that the majority of sources are consistent with their CO(6--5) emission being caused by PDR. However, for the galaxies with r$_{62}$$\gtrsim$$0.24$, it is worth noting that their relatively excited state would require either fairly high star formation rate surface densities ($>$10~M$_\odot$yr$^{-1}$kpc$^{-2}$) or imply the presence of another excitation mechanism (i.e. shocks or XDR). The three most extreme sources in these samples (r$_{62}$$\gtrsim$$0.6$; IRAS~08572$+$3915 at 1.1, NGC~34 at 0.92 and 3C~293 at 0.78) all have a strong indication that AGN activity is responsible for the abnormally high r$_{62}$ \citep[see e.g.][]{Cicone14,Mingozzi18,Emonts05,Floyd06,Papadopoulos10}. Our observed r$_{62}$ limits on J1010$+$0612 and J1100$+$0846 can rule out such an extreme AGN excitation as seen in these sources. Unfortunately our weaker limit on J1430$+$1339, which of the three targets observed in CO(6--5) shows the clearest indications of jet activity \citep{Jarvis19}, does not allow us to place any constraints on the excitation source for the CO(6--5) emission.

In summary, despite the fact that our targets containing kpc-scale ionized outflows \citep[Fig.~\ref{fig:selection};][]{Harrison14, Jarvis19}, we see no evidence that the CO emission is extremely excited based on the \lco(6--5)/CO(2--1) ratios. This result is not entirely unexpected. For example, \citet{Rosenberg15} found that the infrared colours of galaxies is a strong predictor of their CO excitation. Based on this, our galaxies (with \emph{IRAS} 60/100$\mu$m flux $\lesssim$1) should not have highly excited CO. Also, the effect of the AGN is expected to be most clearly seen at J$>$10 \citep[e.g.][]{Mashian15,Lu17} or at extreme gas densities \citep[e.g.][]{Lamperti20}. Observations of higher CO transitions could provide a more complete constraint on the influence of the AGN \citep[see e.g.][]{vanderWerf10,Mashian15,Carniani19} and spatially resolved observations at multiple CO transitions would enable a study of any localised impact on the gas by the AGN or jets which could be undetectable in the total galaxy-wide emission \citep[see e.g.][]{Dasyra16,Zhang19}.

 \subsection{The role of AGN in galaxy evolution}
 \label{sec:galaxy_evolution}
 
 Many works have explored the total molecular gas content of AGN host galaxies compared to non-AGN galaxies \citep[e.g.][]{Simpson12,Husemann17,Kakkad17,Perna18,Rosario18,Shangguan19,Kirkpatrick19}; however, due to the huge amount of variation in the data used, the analysis conducted and the different selection criteria for comparison samples, creating a unified picture of these results is challenging. The most consistent conclusion seems to be that the molecular gas content for low-redshift ($z\ll$1) AGN populations, is broadly consistent with matched non-AGN galaxies \citep[see e.g.][]{Xia12,Krips12,Villar-Martin13}.\footnote{The picture at high redshift is somewhat less clear \citep[e.g.][Circosta et al. in prep.]{Kakkad17,Perna18,Rosario18,Kirkpatrick19}} Our comparison to non-AGN samples generally supports these broad conclusions: we find, at most, moderate differences in observed or derived molecular gas properties for our quasar sample compared to galaxy samples matched in redshift, stellar mass, sSFR and \deltaMS\ (see Fig.~\ref{fig:L'CO_vs_LIR} and Fig.~\ref{fig:derived_qualtities}). Additionally, our results suggest that powerful type 2 AGN with signatures of ionized gas outflows and jets, reside preferentially in gas rich, starburst galaxies.

In Sections \ref{sec:gas_content} and \ref{sec:excitation} we showed that in our sample of local quasars with kpc ionized gas outflows and jets, there is no indication of AGN feedback having an immediate impact on the total gas reservoir once their distance to the star forming main sequence is accounted for. These observations; however, are unable to rule out a more localised impact, which can sometimes be observed using spatially-resolved molecular gas measurements \citep[e.g.][]{Salome17,Rosario18,Fotopoulou19,Ramakrishnan19,Shin19,Lutz20}. Furthermore, we can not rule out that these processes will have an impact on the global molecular gas supply on longer timescales. Specific predictions of the typical spatial scales and time frames of the impact on the molecular gas reservoirs are required to test different AGN feedback models (see e.g.\ \citealt{Lapi14}), which has already started to be investigated on host galaxy star formation rates (see e.g.\ \citealt{Harrison17, Scholtz18, Schulze19}).

Figures \ref{fig:derived_qualtities} and \ref{fig:AGN_comp} indicate that our quasars lie preferentially in molecular gas rich systems even though our only pre-selections were on the width and luminosity of \oiii\ and radio luminosity. Indeed, these systems are more gas rich, and are more likely to reside in starburst galaxies, than less extreme AGN host galaxies (Fig.~\ref{fig:AGN_comp}). This is also in qualitative agreement with recent work revealing a relationship between AGN power and offset from the main sequence \citep[at least at $z\sim$1;][]{Bernhard19,Grimmett20}. Although indirectly, our work is consistent with a link between AGN activity and star formation that is driven by the underlying gas content of the host galaxy. Furthermore, similar results have been found in works considering atomic gas and high redshift sources \citep[see e.g.][respectively]{Ellison19,Rodighiero19}.
 
It is worth noting that one of our sources, which is undetected in our APEX data, may be exceptional in that it does have a low gas content. Using the \citet{Sun14} CO(1--0) luminosity of \lco=1.03$\times$10$^9$~K~\kms~pc$^2$ for J1356$+$1026 would put it amongst the most gas poor sources in our comparison sample from \citet{Tacconi18} (\Mgas/SFR=0.05~Gyr) and cause it to fall $\sim$4 times lower than the \citet{Sargent14} starburst relation (see Fig.~\ref{fig:L'CO_vs_LIR}). This implies either that the luminosity reported in \citet{Sun14} does not detect all of the diffuse, low surface brightness CO emission, or could imply that this source is more rapidly quenched than the rest of our sample. The most obvious exceptional property of this source, which could impact its molecular gas content compared to the rest of the sample, is the double nuclei separated by $\sim$2.5\,kpc \citep{Greene12}, indicating an on-going merger.

Overall the observed high molecular gas masses and incidence of starbursts in our sample are consistent with the scenario where the AGN and star formation are linked, and is in broad agreement with simple evolutionary based AGN unification models \citep[see e.g.][]{Sanders88,Hopkins06,Hickox09}. Specifically, the well studied scenario where gas rich systems have high levels of star formation and obscured / type 2 AGN activity (possibly triggered by mergers) which is followed by feedback processes (such as outflows and jets) that will ultimately quench the AGN activity and star formation in the galaxy. Larger, less biased samples would be needed to confirm these models however. Although we can not be sure of the fate of our galaxies, we may have caught these systems in a special evolutionary phase where the feedback processes are just beginning. We can concretely conclude that the outflows and jets we observe do not {\em rapidly} remove the global molecular gas in an appreciable way (i.e.\ on a timescale shorter than, or equal to, the observed quasars, jets or outflows).

Our findings are consistent with many previous studies of the molecular gas and star formation in low redshift AGN ($z$$\lesssim$$0.2$). For example, \citet{Husemann17} find gradually increasing amounts of molecular gas going from AGN in bulge dominated to disc dominated to major merger host galaxies and a trend to higher molecular gas masses in systems with more luminous AGN. Similarly, \citet{Bertram07} find that the Seyferts in their sample have molecular gas content consistent with normal star forming galaxies, while the powerful quasars are more consistent with starbursts. Luminous AGN are known to generally reside in galaxies with more recent star formation than their lower luminosity counterparts \citep[see e.g.][]{Balmaverde16,Bernhard19,Grimmett20,Kim20}. Finally, there is evidence that obscured AGN lie in more gas rich systems than their un-obscured counterparts \citep{Wylezalek16,Rosario18} and that the most extreme outflows may be preferentially found in rapidly star-forming, gas rich systems \citep[see e.g.][]{RodriguezZaurin13,Harrison14,Wylezalek16}.

To summarize, we find that our sample, selected to be luminous type 2 AGN hosting ionized outflows, lie preferentially in gas rich galaxies, with high levels of simultaneous star formation, which is consistent with the evolutionary framework described above. However, the small size of this sample and the two non-detections limit our ability to expand these findings to the quasar population in general. By selecting systems with fast, prominent kpc ionized gas outflows we might have expected these outflows to be able to remove the molecular gas, resulting in a deficit. However, the data suggest that if these outflows or jets will ultimately have an impact on the global molecular gas content, it is subtle, or we have captured them too early in the feedback process for this effect to be measurable.  

\section{Conclusions}
\label{sec:conclusions}

Using APEX observations of the CO(2--1) emission line we have explored the global molecular gas content of nine $z$~$\sim$~0.1 galaxies selected to host powerful type 2 quasars ($\log [L_{\text{AGN}}$/erg\,s$^{-1}]$$\gtrsim$$45$) with galaxy-wide ionized outflows and radio jets \citep[see Fig.~\ref{fig:selection};][]{Harrison14,Jarvis19}. We detected seven of the nine targets in CO(2--1), with corresponding \lco(2--1) values of (1.4--7)$\times$10$^9$~K~km~s$^{-1}$ pc$^2$. For a subset of three targets we used APEX to obtain upper limits on the CO(6--5)/CO(2--1) emission-line ratios. Our main conclusions are: 

\begin{enumerate}

    \item For at least seven of the nine quasars in our sample, the total molecular gas reservoirs show no indication of being rapidly depleted due to AGN feedback, despite being selected to have powerful ionized gas outflows and jets. Firstly, we find CO luminosities consistent (within 0.3~dex) with what would be predicted for the general galaxy population given their \lir\ and distance to the star-forming main sequence (see Fig.~\ref{fig:L'CO_vs_LIR} and Section~\ref{sec:lco_vs_lir}).
    Secondly, the derived gas fractions and depletion times of our seven CO(2--1) detected sources (i.e. \Mgas/M$_\star$$\approx$$0.1$--1.2 and \Mgas/SFR$\approx$0.16--0.95~Gyr, respectively) are comparable to those of redshift-matched non-AGN star-forming galaxies when taking into account their stellar mass, specific star formation rate and distance from the main sequence (see Fig.~\ref{fig:derived_qualtities} and Section \ref{sec:derived_quantities}).

   \item Galaxies hosting powerful AGN (i.e. log(\oiii/H$\beta$)$\gtrsim$0.6) tend to have systematically higher gas fractions than those with less powerful AGN and star-forming galaxies in general, when our sample is considered together with those from the xCOLD GASS survey \citep{Saintonge17}. Galaxies across these samples with the highest gas fractions appear to contain the most powerful AGN and highest levels of concurrent star formation (in relation to the star-forming main sequence; see Fig.~\ref{fig:AGN_comp} and Section~\ref{sec:AGN}).

    \item The AGN are not having an extreme impact on the global CO excitation in at least two of the three sources for which we have upper limits on the \lco(2--1)/CO(6--5) emission-line ratios (i.e.\ r$_{62}$ $\lesssim$ $0.66$; see Fig.~\ref{fig:excitation} and Section~\ref{sec:excitation}).

\end{enumerate}

In summary, we find that the majority of our sample of quasars have gas rich, starburst host galaxies, even though we did not select the sample based on these properties. Furthermore, we find that their gas masses are consistent with what would be expected for their observed levels of star formation. There are no signs of an instantaneous depletion of the total molecular gas reservoir by the AGN in our sample, despite their high bolometric luminosities, strong ionized gas outflows and the presence of kpc scale jets in many. Our results are consistent with a requirement for high molecular gas fractions to feed both quasar activity and intense periods of star formation. Indeed, by selecting luminous AGN with powerful ionized gas outflows, we may have predominantly selected galaxies in a phase in their evolution where intense star formation and AGN activity are powered by large molecular gas reservoirs and the ``feedback'' in the form of jets and outflows is relatively young and these processes have not yet had any global impact upon the host galaxies. 

Future, higher resolution CO observations and observations of more CO transitions will help determine if these processes have a more subtle and / or localised impact upon the molecular gas properties. Furthermore, galaxy formation models should work towards specific predictions of the molecular gas properties (e.g.\ gas fractions, depletion times, excitation) to compare to observations, such as ours, to aid understanding of the expected physical scales and time frames of any impact caused by different AGN feedback model prescriptions.  

\section*{Acknowledgements}
 We thank the referee for their prompt and constructive comments. APEX is a collaboration between the Max-Planck-Institut f\"ur Radioastronomie, the European Southern Observatory, and the Onsala Observatory. PJ acknowledges funding from the European Research Council (ERC) under the European Union's Horizon 2020 research and innovation programme under grant agreement No 724857 (Consolidator Grant ArcheoDyn).

\section*{Data Availability Statement}
 The data underlying this article were accessed from the ESO Science Archive Facility: \url{http://archive.eso.org/eso/eso_archive_main.html} under programme IDs E-0100.B-0166 and E-0104.B-0292. The derived data generated in this research will be shared on reasonable request to the corresponding author.




\bibliographystyle{mnras} %



\appendix

\section{Bayesian fitting}
\label{app:bayesian}

This section provides further details about our Bayesian fits to our APEX CO observations. The values quoted in Table~\ref{tab:apex_fits} are derived from these fits. The corner plots showing the posterior probability distributions of each of the parameters for each source are given in the online supplementary data (`Supplement to Appendix A'; Fig.~\ref{fig:app:J0945+1737_corner}-\ref{fig:app:J1430+1339_corner_6-5}).

For the seven CO(2--1) detected targets (see Section~\ref{sec:results}), we used initial guess parameters from reduced $\chi^2$ Gaussian fits to the emission-line data using 100~km~s$^{-1}$ bins. For the initial guess parameters for the two CO(2--1) undetected targets (J0958$+$1439 and J1356$+$1026) we used the average $\sigma$ (line width) from the detected targets (170~\kms) and $v_p$=0. For J0958$+$1439 we chose an initial guess flux derived from the \lco\ -- \lir\ starburst relation (see Section \ref{sec:lco_vs_lir}; f$\approx$6~Jy \kms) and for J1356$+$1026 we used the \citet{Sun14} ALMA CO(1--0) and CO(3--2) total fluxes for a rough estimate (f$\approx$6~Jy \kms). For the initial guess parameters for fitting the CO(6--5) data we used the values found through our Bayesian analysis for the CO(2--1) data, multiplying the fluxes by 1.4 to convert to the CO(6--5) transition \citep[for typical LIRGs;][]{Papadopoulos12}. By fitting sources multiple times with the initial guesses varied by approximately an order of magnitude, we confirmed that the results, within errors do not depend strongly on the initial guess used.

We adopted weak priors for our fitting procedure. We limited the flux and $\sigma_N$ (noise) to be greater than zero, $v_p$ to be within $+$/$-$2000~km/s for CO(2--1) and $+$/$-$1500~km/s for the CO(6--5) data (i.e.\ the velocity coverage of the data). We constrained $\sigma$ (line width) to be greater than zero and used slightly different maximum values of $\sigma$ for different cases. Specifically, for the CO(2--1) detections we limited the line width to be less than 3 times the width of the initial guess from the reduced $\chi^2$ Gaussian fit (corresponding to upper values of 360--680~\kms), and for the non-detections we used the largest limit from the detected lines (i.e.\ $\sigma \leq$680~\kms). For the CO(6--5) data we limited $\sigma$ (line width) to be less than 3 times the CO(2--1) line width from this Bayesian analysis (i.e.\ the values quoted in Table \ref{tab:apex_fits}). We note that using more complicated or more constraining priors could lower the errors on our fits and our upper limits; however this would risk introducing bias into the results.

Our fitting code is designed to be completely general and therefore, our likelihood is composed of a single Gaussian with both Poisson and Gaussian noise considered. However, Poisson and Gaussian likelihoods become indistinguishable even for very moderate values of the mean parameter (>10). Our priors are uniform distributions which drop to 0 outside of the bounds described above. We use 100 walkers for the MCMC analysis and run it for 500 steps with the first 300 steps later cropped as burn in (visual inspection of the trace plots were used to confirm the burn in period for all fits). We determine the starting position for each walker using the initial guesses described above with a random offset added drawn from a uniform distribution limited to within $\pm$ 3 orders of magnitude less than the initial guess. 

For each source where the CO(2--1) line is detected (all except J0958$+$1439 and J1356$+$1026) the posterior distributions for all 4 parameters show clear peaks. In contrast, the non detections do not show clear peaks in one or more of the parameters (the posteriors for $\sigma$ in particular do not have a clear peaks for any of the non detected CO lines; see corner plots in online `Supplement to Appendix A'). Further details on the individual fits can be found in the corner plots and captions in Figures~\ref{fig:app:J0945+1737_corner}-\ref{fig:app:J1430+1339_corner_6-5}.

 \section{\texorpdfstring{\alphaco}{alpha CO} and molecular gas mass calculations}
\label{app:alphaca}
 
Here we provide specific details about how we calculated \alphaco\ and the molecular gas masses. We also present these derived values for each source, using the two different methods that are discussed in this paper (see Table \ref{tab:app:Mgas}). In each case r$_{21}=0.8$ is used to convert from CO(2--1) to CO(1--0) (see Section~\ref{sec:derived_quantities}).

For a comparison with the values from \citet{Tacconi18} (Section~\ref{sec:derived_quantities} and Fig.~\ref{fig:derived_qualtities}) we follow the \alphaco\ calculation from that work \citep[as in][]{Genzel15}, taking the geometric mean of the metallicity dependent \alphaco\ recipes of \citet{Bolatto13} and \citet{Genzel12}: 
\begin{equation}
\alpha_\textrm{CO} = 4.36 \times \sqrt{ \begin{aligned} 0.67 \times \exp(0.36 \times 10^{-1\times(12+\log\textrm{(O/H)}-8.67)} \\ \times 10^{(-1.27 \times (12+\log\textrm{(O/H)}-8.67)} ) \end{aligned} } ,  
\end{equation}
where \alphaco\ has units M$_\odot$ ( K km s$^{-1}$ pc$^2$)$^{-1}$. Also following \citet{Tacconi18} we use the following mass metallicity relation from \citet{Genzel15}:
\begin{equation}
    12+\log\textrm{(O/H)} = a - 0.087 \times (\log M_\star - b)^2,
\end{equation}                                                  where $a$=8.74 and
\begin{equation}
b=10.4+4.46 \times \log(1+z)-1.78 \times (\log(1+z))^2.  
 \end{equation}                                               
 
For the comparison with the xCOLD GASS samples in Section \ref{sec:AGN} we calculate \alphaco\ following \citet{Saintonge17}. Specifically, they use the metallicity and \deltaMS\ dependent \alphaco\ correlation from \citet{Accurso17}: 
 \begin{equation}
     \log \alpha_\textrm{CO}=14.752-1.623\times [12+\log(\textrm{(O/H)}] + 0.062 \times \log \Delta_\textrm{MS}. 
 \end{equation}
 To keep consistent with the methods adopted in the comparison sample, we calculate the metallicity (12+log(O/H)) following the \citet{Pettini04} `O3N2' consistent mass--metallicity relation of \citet{Kewley08}: 
 \begin{equation}
     12+\log\textrm{(O/H)}=a+b \times \log M_\star+c\times \log M_\star ^2+d\times\log M_\star^3, 
 \end{equation}
 where a=32.1488, b=$-$8.51258, c=0.976384, d=$-$0.0359763 and $M_\star$ is in M$_\odot$. For additional consistency we do not use the \deltaMS\ (i.e.\ the ratio of the sSFR of the galaxy and its local main sequence) derived in the main paper, but re-calculate this value for use in the $\alpha_\textrm{CO}$ calculation, using the same method as in \citet{Saintonge17}. That is, using the star forming main sequence from \citet{Accurso17}: 
\begin{multline}
    \log \textrm{sSFR}_\textrm{MS} [\textrm{Gyr}] = -1.12 + 1.14\times z-0.19\times z^2-(0.3 + 0.13\times z) \\ \times (\log M_\star -10.5) .       
\end{multline}

As described in \cite{Accurso17}, this \alphaco\ relation is only valid within 7.9$<$12+log(O/H)$<$8.8 and $-$0.8$<$log\deltaMS$<$1.3. Using the \deltaMS\ values for these calculations J0945$+$1737, J1000$+$1242 and J1356$+$1026 all fall outside (or at the edge of) of the allowed \deltaMS\ range (with log \deltaMS=1.6, 1.4 and 1.3 respectively). However using the recommended starburst \alphaco\ from \citet{Accurso17} of 1 for these sources does not change the conclusions of this work.

We note that the gas masses derived from both of the methods described above are consistent within errors.


\begin{table}
 \caption{The values \alphaco\ and \Mgas\ for each source in this sample using the two methods used in this work to be consistent with the literature comparisons.
 \newline Notes: (1) Object name; (2) \alphaco\ calculated to be consistent with \citet{Tacconi18}; (3) log \Mgas\ calculated to be consistent with \citet{Tacconi18}; (4) \alphaco\ calculated to be consistent with xCOLD GASS \citep{Saintonge17}; (5) log \Mgas\ calculated to be consistent with xCOLD GASS \citep{Saintonge17}.}

	\begin{tabular}{c|cc|cc} 
	\hline
&\multicolumn{2}{c}{Tacconi et al.}&    \multicolumn{2}{c}{xCOLD GASS} \\
&\multicolumn{2}{c}{(Fig.\ref{fig:derived_qualtities})}&    \multicolumn{2}{c}{ (Fig.\ref{fig:AGN_comp}) } \\
	\hline   
Name & \alphaco & log \Mgas / M$_\odot$ & \alphaco & log \Mgas / M$_\odot$\\
 (1) & (2) & (3) & (4) & (5) \\
	\hline   
J0945+1737 & 4.1 & 10.07$^{+0.1}_{-0.08}$ & 5.3 & 10.18$^{+0.1}_{-0.08}$ \\
J0958+1439 & 4.0 & $<$10.0 & 3.4 & $<$10.0 \\
J1000+1242 & 4.2 & 9.99$^{+0.08}_{-0.06}$ & 6.0 & 10.14$^{+0.08}_{-0.06}$ \\
J1010+1413 & 4.1 & 10.54$^{+0.07}_{-0.06}$ & 3.7 & 10.5$^{+0.07}_{-0.06}$ \\
J1010+0612 & 4.0 & 10.03$^{+0.1}_{-0.09}$ & 3.9 & 10.01$^{+0.1}_{-0.09}$ \\
J1100+0846 & 4.0 & 10.15$^{+0.08}_{-0.06}$ & 3.7 & 10.1$^{+0.08}_{-0.06}$ \\
J1316+1753 & 4.1 & 10.1$\pm$0.2 & 3.7 & 10.1$\pm$0.2 \\
J1356+1026 & 4.0 & $<$10.0 & 3.9 & $<$10.0 \\
J1430+1339 & 4.1 & 9.9$^{+0.2}_{-0.1}$ & 3.3 & 9.8$^{+0.2}_{-0.1}$ \\
	\hline   
	\end{tabular}
    
\label{tab:app:Mgas} 

	\end{table}

\setcounter{section}{1}

\refstepcounter{figure}
  \label{fig:app:J0945+1737_corner}

 \refstepcounter{figure}
  \label{fig:app:J0958+1439_corner}

 \refstepcounter{figure}
  \label{fig:app:J1000+1242_corner}
 
 \refstepcounter{figure}
  \label{fig:app:J1010+1413_corner}
 
 \refstepcounter{figure}
  \label{fig:app:J1010+0612_corner}
 
 \refstepcounter{figure}
  \label{fig:app:J1100+0846_corner}

\refstepcounter{figure}
  \label{fig:app:J1316+1753_corner}
 
 \refstepcounter{figure}
  \label{fig:app:J1356+1026_corner}

\refstepcounter{figure}
  \label{fig:app:J1430+1339_corner}
  
\refstepcounter{figure}
  \label{fig:app:J1010+0612_corner_6-5}
 
 \refstepcounter{figure}
  \label{fig:app:J1100+0846_corner_6-5}

\refstepcounter{figure}
  \label{fig:app:J1430+1339_corner_6-5}


\bsp	
\label{lastpage}
\end{document}